\documentclass[preprint, 3p]{elsarticle}

\usepackage{lineno,hyperref}
\modulolinenumbers[5]

\journal{JSV}

\usepackage[usenames,dvipsnames]{xcolor}

\usepackage{graphicx}        

\usepackage{natbib}

\usepackage{amsmath}
\usepackage{amsfonts}
\usepackage{amssymb}

\usepackage{subfigure}
\usepackage{xspace}
\newcommand{\bs}{\boldsymbol}

\newcommand{\mbf}{\mathbf}

\newcommand{\ie}{i.\,e.\xspace}
\newcommand{\eg}{e.\,g.\xspace}
\newcommand{\cf}{c.\,f.\xspace}
\newcommand{\fp}{\,.}
\newcommand{\fk}{\,,}

\newcommand{\mrm}[1]{\mathrm{#1}} 

\newcommand{\nc}{\newcommand}
\newcommand{\rnc}{\renewcommand}
\nc{\VINES}{vibro-impact NES\xspace}
\nc{\sref}[1]{Sect.~\ref{sec:#1}}
\nc{\srefs}[1]{Sect.~\ref{sec:#1}}
\nc{\srefo}[1]{\ref{sec:#1}}
\nc{\eref}[1]{Eq.~\ref{eq:#1}}
\nc{\erefs}[1]{Eqs.~\ref{eq:#1}}
\nc{\erefo}[1]{\ref{eq:#1}}
\nc{\fref}[1]{Fig.~\ref{fig:#1}}
\nc{\frefs}[1]{Figs.~\ref{fig:#1}}
\nc{\frefo}[1]{\ref{fig:#1}}
\nc{\tref}[1]{Tab.~\ref{tab:#1}}
\nc{\e}[2]{\begin{equation}\label{eq:#2}#1\end{equation}}
\nc{\ea}[2]{\begin{eqnarray}\label{eq:#2}#1\end{eqnarray}}
\rnc{\matrix}[2]{\left[\!\!\begin{array}{#1}
	#2\end{array}\!\!\right]}
\rnc{\vector}[1]{\matrix{c}{#1}}
\nc{\dd}{\mathrm{d}}
\usepackage{color}
\nc{\COMMENT}[1]{\textcolor{red}{#1}}
\nc{\COMMENTth}[1]{\textcolor{blue}{#1}}
\usepackage{wasysym}
\usepackage{tikz} 
\nc{\circled}[1]{(#1)}

\usepackage{stfloats}
\fnbelowfloat 

\makeindex             

\begin{document}

\begin{frontmatter}
\title{Effects of modal energy scattering and friction on the resonance mitigation with an impact absorber}
\author[addressILA]{Timo Theurich}
\author[addressILA]{Johann Gross}
\author[addressILA]{Malte Krack}

\address[addressILA]{Institute of Aircraft Propulsion Systems, University of Stuttgart, Pfaffenwaldring 6, 70569 Stuttgart, Germany\\ theurich@ila.uni-stuttgart.de, gross@ila.uni-stuttgart.de, krack@ila.uni-stuttgart.de}

\begin{abstract}
A linear vibration absorber can be tuned to effectively suppress the resonance of a particular vibration mode.
It relies on the targeted energy transfer into the absorber within a narrow and fixed frequency band.
Nonlinear energy sinks (NES) have a similar working principle.
They are effective in a much wider frequency band but generally only in a limited range of excitation levels.
To design NES, their working principle must be thoroughly understood.
We consider a particular type of NES, a small mass undergoing impacts and dry friction within a cavity of a base structure (vibro-impact NES or impact absorber).
The nonlinear dynamic regimes under near-resonant forcing  and resulting operating ranges are first revisited.
We then investigate how off-resonant vibration modes and dissipation via impacts and dry friction contribute to the vibration suppression.
Moreover, we assess the effectiveness of the impact absorber for suppressing multiple resonances in comparison to a linear tuned vibration absorber (LTVA) and a pure friction damper with the same mass.
\end{abstract}

\begin{keyword}
Nonlinear energy sink \sep targeted energy transfer \sep strongly modulated response \sep energy scattering \sep friction damper \sep linear tuned vibration absorber
\end{keyword}

\end{frontmatter}

\section{Introduction} \label{sec:intro}
This work addresses the passive vibration mitigation of systems under \emph{harmonic near-resonant forcing} (rather than impulsive or self-excitation).
A common means to mitigate resonances is the linear tuned vibration absorber (LTVA) or tuned mass damper.
Here, a small mass is elastically attached to the base or \emph{primary structure}.
We call such a secondary structure \emph{absorber}, if its purpose is to mitigate the vibrations of the directly excited primary structure, regardless of its working principle (dissipation, localization, dispersion, ...).
The working principle of the LTVA relies on \emph{targeted energy transfer} and \emph{localization}: At the natural frequency of the absorber, the vibration energy is effectively \emph{pumped} from the primary into the secondary structure.
As the vibration level of the primary structure is reduced, the external forcing can supply less power to the whole system.
In the theoretical case of an undamped absorber, the steady-state vibration level of the primary structure goes to zero at the absorber's natural frequency.
This frequency is commonly tuned to a particular resonance frequency of the primary structure.
Near this frequency, two resonances occur, one at a slightly lower and one at a slightly higher frequency.
The LTVA's damping can be designed to achieve equal levels of these two resonance peaks (Den-Hartog's equal peak method \cite{DenHartog1928}).
This leads to a considerable reduction of the vibration response around the specific resonance frequency of the primary structure without absorber.
\\
The main weakness of the LTVA is that it only operates effectively near its natural frequency.
If its natural frequency is fixed, the LTVA has virtually no effect on resonances other than the one it is tuned to.
A detuning also occurs if the targeted resonance frequency shifts during operation of the system (\eg due to stress stiffening, loading/unloading of mass).
A large variety of means of \emph{increasing the operating frequency bandwidth} have been proposed.
In what follows, we mention only passive means, \ie, those that do not rely on electrical power for sensing or actuation.
\\
For the specific case of rotating machinery, the \emph{centrifugal pendulum vibration absorber} was applied with considerable success to mitigate torsional vibrations in automotive and aircraft engines \cite{Olson2005}.
Here, the additional mass is driven on a curved, typically circular path.
Its natural frequency therefore depends on the centripetal acceleration and thus on the rotor speed.
The path can be designed so that the absorber frequency follows a specific multiple of the rotor speed, which is particularly interesting for counteracting engine-order excitations.
It should be emphasized that the working principle of this type of absorber can be explained by linear vibration theory.
\\
\emph{Self-adapting systems} have the potential to passively adjust their configuration and, thus, their dynamic behavior to an external excitation.
Such systems may consist of a beam or a string on which a mass can slide.
The position of the mass determines the natural frequencies of the system.
In \cite{Babitsky1993,Thomsen1996,Miller2013,Krack2016}, self-damping and self-resonating systems are investigated, where the system re-arranges itself depending on the excitation frequency to achieve and maintain a state of either low or high vibration level, respectively.
A self-resonating system could be used as an absorber, with the interesting capability that the absorber passively tunes its operating frequency to the excitation frequency, and thus broadens the operating frequency bandwidth.
\\
The \emph{nonlinear tuned vibration absorber} introduced in \cite{Habib2015} can be seen as an extension of the LTVA to nonlinear primary structures.
At higher vibration levels, such structures commonly show stiffening or softening behavior, \ie, an in-/decrease of the effective natural frequency.
Hence, conventional LTVAs may not perform well for all vibration levels.
By properly tailoring the functional form of the absorber's nonlinearity in accordance with the primary structure's nonlinearity, the resonant vibrations can be effectively mitigated over a wide range of excitation levels.
\\
If one replaces the, at least for sufficiently small vibration levels, linear attachment of the LTVA by an essentially nonlinear one, the absorber is referred to as \emph{nonlinear energy sink} (NES) \cite{Vakakis2008}.
They can generally be used to mitigate the response under impulsive or harmonic loading as well as self-excitation \cite{Lee2007I,Lee2007II,Gendelman2008,Hubbard2014}.
For harmonic loading, it has been shown that a NES can mitigate the resonances with many modes of the primary structure\footnote{In the context of NES, the primary structure is often referred to as \emph{linear oscillator}.}, without the need to adjust the absorber properties \cite{Vakakis2008,Gendelman2008}.
However, such absorbers generally operate effectively only in a limited range of vibration levels.
Like in the case of the LTVA, the main working principle of NES is targeted energy transfer from the primary structure to the secondary structure.
The essential nonlinearity leads to intricate dynamic regimes and nonlinear mechanisms of the primary-to-secondary-structure energy transfer, some of which will be revisited later in this article.
Moreover, the nonlinearity opens the door for energy transfer from low to high frequency modes within the primary structure itself.
Assuming the same damping ratio for all modes of the linearized system, the energy of high-frequency modes dissipates faster.
This permits a quicker decay of the impulse response as \eg demonstrated experimentally and numerically in \cite{Al-Shudeifat2013,Luo2014,Al-Shudeifat2015}.
In contrast, for the forced response under sustained near-resonant loading, it is commonly found that the off-resonant modes' steady-state response level is at least one order of magnitude smaller than that of the resonant mode \cite{Starosvetsky2008,Zulli2014,Luongo2015}. Thus, it is concluded that off-resonant modes are not relevant to explain the working principle of NES under harmonic loading.
It should be emphasized that only smooth nonlinearities, namely cubic springs were considered in \cite{Starosvetsky2008,Zulli2014,Luongo2015}.
\\
In by far most theoretical works on NES, a cubic spring is considered as essentially nonlinear attachment of the absorber.
An alternative essential nonlinearity is unilateral contact, which can be easily realized by designing the additional mass to undergo impacts with the primary structure.
This form of NES is commonly referred to as vibro-impact NES.
The effectiveness of \VINES under harmonic loading is well-supported by analytical, numerical and experimental results, see \eg \cite{Gourc2015}.
The perhaps most mature use of \VINES for resonance mitigation is their industrial application to turbomachinery blades and vanes \cite{Hartung2004,Hartung2016}.
Here, a spherical additional mass is placed into a cavity of the primary structure.
Besides the usual working principle of NES, dissipative impacts of \VINES could also contribute to the vibration mitigation.
In fact, the setup of \emph{impact dampers} is generally the same as that of \VINES \cite{Masri1973,Brown1988}.
Along similar lines, it has been proposed to design centrifugal pendulum absorbers with a limited moving space in order to provide additional damping by dissipative impacts at high vibration levels \cite{Duffy2000,Shaw2006}.
For impact dampers, it is typically concluded that they operate most effectively over a large frequency band for relatively large additional masses and strongly dissipative impacts.
In contrast, it was demonstrated by numerical simulation in \cite{Hartung2016} that a considerable reduction of the resonant response level is achieved even in the case of a conservative impact model, and relatively small absorber mass (significantly less than $1\%$ of the primary structure).
From a practical point of view, this is an extremely important finding, since the immediate and local \emph{dissipation by impacts} is inevitably associated with plastic deformation and, thus, most likely \emph{limits the service life} of absorber or primary structure.
\\
The purpose of the present work is to better understand the working principle of impact absorbers for mitigating resonant vibrations.
We will see that in some operating ranges, these devices mainly rely on damping, while in other ranges, their effectiveness can be explained by the theory of nonlinear energy sinks and targeted energy transfer.
Hence, we deliberately avoid the terms \VINES and impact damper, and will refer to these devices as \emph{impact absorbers}.
In \sref{modsim}, we present the model considered for the numerical investigations throughout this work.
We demonstrate that the model is qualified to reproduce the well-known dynamic regimes and operating ranges of impact absorbers in \sref{modquali}.
We analyze how off-resonant modes contribute to their working principle in \sref{scattering}.
The effect of dissipation by dry friction and impacts on the system dynamics and the absorber performance is addressed in \sref{dissipation}.
Finally, we assess the performance of the impact absorber in comparison to a conventional LTVA and a pure friction damper in \sref{comparison}.
This article ends with conclusions.

\section{Model and simulation methods}\label{sec:modsim}
We consider the model illustrated in \fref{model}.
The primary structure is a cantilevered beam undergoing bending vibrations in the depicted plane.
The absorber is a lightweight mass permitted to undergo dry friction and impacts with the cavity walls.
The higher effectiveness of impact absorbers for an increasing mass ratio $\varepsilon=\frac{m_\mrm{a}}{m_\mrm{s}}$ of the absorber mass $m_\mrm{a}$ to the mass of the primary structure $m_\mrm{s}$ is well known and has been demonstrated in various studies (see \eg \cite{Hartung2016},\cite{Masri1973}). The influence of this ratio on the findings of this work is not of interest here and thus it is kept constant at $\varepsilon=2\%$.
\begin{figure}[b!]
	\centering
	\includegraphics[width=1\textwidth]{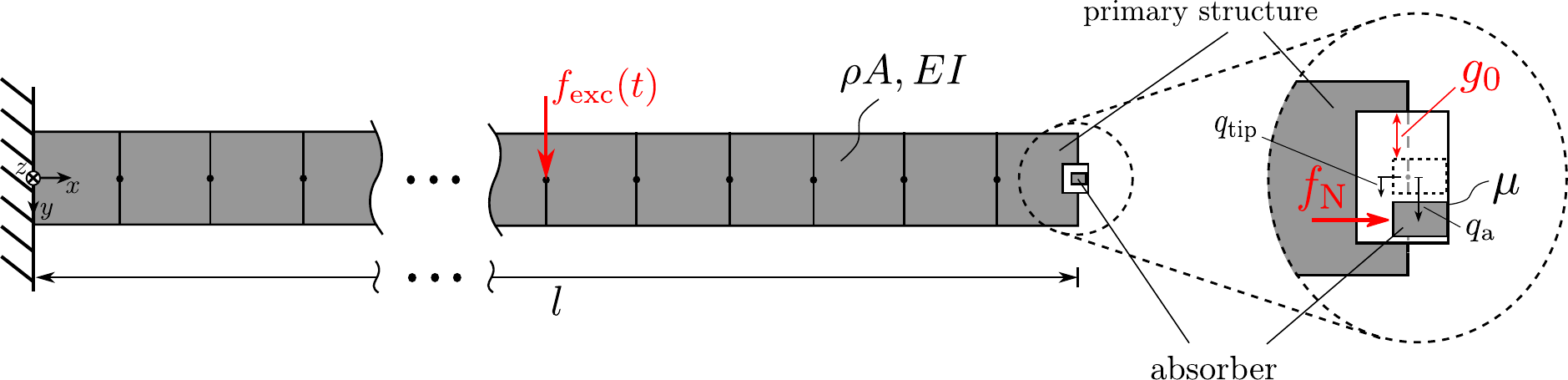}
	\caption{Finite Element model of a cantilevered beam undergoing impacts and dry friction with a lightweight absorber.}
	\label{fig:model}
\end{figure}
\begin{table}[b!]
	\centering
	\caption{Nominal model parameters}
	\begin{tabular}{|c|c|c|c|c|c|c|c|}
		\hline
		 Quantity & $N_\mrm{e}$ & $\rho A$ & $EI$ & $l$ & $g_0$ & $e$ & $\varepsilon=\frac{m_\mrm{a}}{m_\mrm{s}}=\frac{m_\mrm{a}}{\rho A l}$\\
		\hline
		Value & 30 & 1.1275 & 362.88 & 0.5 & 0.001 & 1 & 2\% \\
		\hline
	\end{tabular} \label{tab:model}
\end{table}

\subsection{Primary structure model}
The beam of length $l$ is modeled in accordance with linear Euler-Bernoulli theory (flexural rigidity $EI$, mass per unit length $\rho A$).
The reduced mass and stiffness due to the cavity is neglected in the beam model.
$N_\mrm{e}=30$ finite elements are used for spatial discretization.
Each node has one transversal ($y$-direction) and one rotational (about $z$-axis) degree of freedom.
Hermitian polynomials are used as shape functions.
The differential equation of motion of the free beam, \ie, \emph{without contact} can be stated as
\e{
\mbf M_\mrm{s} \ddot{\bs q}_\mrm{s}(t) + \mbf C_\mrm{s} \dot{\bs q}_\mrm{s}(t) + \mbf K_\mrm{s} \bs q_\mrm{s}(t) = \bs p(t)\fk
}{eqmbeam}
where $\bs q_\mrm{s}$ are the nodal coordinates of the finite element model, overdots denote differentiation with respect to time $t$ and $\mbf M_\mrm{s}$, $\mbf C_\mrm{s}$, $\mbf K_\mrm{s}$ are mass, viscous damping and stiffness matrices, respectively.
The external forcing $\bs p(t)$ is applied as concentrated load in the transverse direction at $x=0.4l$, which corresponds to the 12-th node (\cf \fref{model}).
A harmonic forcing
\e{f_{\mathrm{exc}}(t) = \hat{f}_\mrm{exc} \cos\Omega t}{exc}
with magnitude $\hat{f}_\mrm{exc}$ and angular frequency $\Omega$ is considered.
\\
A modal truncation,
\e{ \bs q_\mrm{s}(t) \approx
\sum\limits_{i=1}^{N_\mrm{m}} \bs\varphi_i \eta_i
\fk
}{modalred}
to the $N_{\mathrm m}$ lowest-frequency modes of the free and undamped beam is used.
Herein, $\bs\varphi_i$ are the mass-normalized modal deflection shapes and $\eta_i$ are the associated modal coordinates.
The truncation satisfies two purposes:
First, it permits to deliberately suppress off-resonant modes and thus facilitates the systematic investigation of the energy transfer to and dissipation by off-resonant modes.
Second, it allows larger time step sizes in the explicit numerical integration, as detailed later, allowing more thorough parameter studies under limited computational resources.
By projecting the dynamics onto the truncated set of modes, \eref{eqmbeam} is replaced by
\e{
\ddot{\eta_i}(t) + 2D_i\omega_i\dot{\eta}_i + \omega_i^2\eta_i = \bs \varphi_i^\mrm{T} \bs{f}_\mrm{exc} \qquad i = 1,\ldots,N_\mrm{m}\fp
}{eqmmod}
Herein, $\omega_i$, $D_i$ are the modal frequency and damping ratio.
Unless otherwise stated, a modal damping ratio of $D_i=1\%$ is specified for all considered modes.

\subsection{Absorber and contact model}
The absorber is modeled as a rigid body with only one transverse degree of freedom, $q_\mrm{a}$ ($y$-direction).
\emph{Without contact} and neglecting gravitational acceleration, the absorber's equation of motion thus simply reads
\e{
m_\mrm{a}\ddot{q}_\mrm{a}(t)=0\fk
}{eqmabs}
where $m_{\mathrm{a}}$ is the absorber mass.
\\
Contact interactions are taken into account between the absorber and the beam's tip node (\cf \fref{model}).
First, the absorber is not permitted to penetrate the upper and lower cavity walls.
$q_\mrm{a}$ counts from $y = 0$, \ie, the middle of the cavity.
If $\left(q_\mrm{a}-q_\mrm{tip}\right) = g_0$, where $g_0$ is the nominal clearance between absorber and either cavity wall, then absorber and beam are in contact at the lower wall.
Then, a compressive reaction force, $\lambda_{\mrm{u,}1}\geq 0$, acts between the bodies that is large enough to avoid penetration.
If $\left(q_\mrm{a}-q_\mrm{tip}\right)<g_0$, the contact at the lower cavity wall is open and $\lambda_{\mrm{u,}1}=0$.
Similarly, $-\left(q_\mrm{a}-q_\mrm{tip}\right)\leq g_0$ has to hold to avoid penetration with the upper cavity wall, and $\lambda_{\mrm{u,}2}\geq 0$ for the associated reaction force $\lambda_{\mrm{u,}2}$.
The above described complementary inequalities are referred to as the \emph{Signorini law}. These constraints can be summarized as
\ea{
	& 0\leq g_1\left(\bs q_\mrm{s},q_\mrm{a}\right)  \perp  \lambda_{\mrm{u,}1}\geq 0 \qquad & g_1 = +\left(q_\mrm{a}-q_\mrm{tip}\right)+g_0 \\
	& 0\leq g_2\left(\bs q_\mrm{s},q_\mrm{a}\right)  \perp  \lambda_{\mrm{u,}2}\geq 0 \qquad & g_2 = -\left(q_\mrm{a}-q_\mrm{tip}\right)+g_0\fp
}{signorini}
Herein, $\perp$ indicates the complementary character of the problem: If the contact gap $g_i$ is open, the associated contact force is zero, and if the force is positive, the contact gap is closed.
\\
Friction between absorber and primary structure occurs if there is a nonzero normal load $f_{\mathrm N}$ (\cf \fref{model}).
The normal preload could be caused by gravity forces.
We also have in mind the specific application to rotating turbomachinery blades where the normal load is mainly caused by centrifugal forces.
It was shown in \cite{Mansour1974} for a similar setup that dry friction can affect the stability of certain dynamic regimes.
If frictional dissipation turns out to be beneficial in certain operating ranges, there are many ways to deliberately introduce the required preload.
It is assumed that the contact with the right cavity wall is always closed and the normal load is considered as constant.
In the theoretical case where the gap is sufficiently large to avoid contacts with the upper and lower cavity walls, the absorber acts as a pure friction damper.
This greatly facilitates the comparative performance assessment between impact absorbers and friction dampers in \sref{comparison}.
\\
The dry friction is modeled by the \emph{Coulomb law} with friction coefficient $\mu$.
It can be summarized as
\begin{equation} \label{eq:coulomb}
\begin{cases}
\gamma=0 \quad \Leftrightarrow \quad |\lambda_{\mrm{f}}|<\mu f_\mrm{N} &\quad \text{absorber sticking}\\
\gamma>0 \quad \Leftrightarrow \quad \lambda_{\mrm{f}}=-\mu f_\mrm{N} &\quad \text{absorber sliding down}\\
\gamma<0 \quad \Leftrightarrow \quad \lambda_{\mrm{f}}=\mu f_\mrm{N} &\quad \text{absorber sliding up}\\
\end{cases}.
\end{equation}
Herein, $\gamma=\dot q_\mrm{a}-\dot q_\mrm{tip}=\dot g_1=-\dot g_2$ is the relative velocity between absorber and beam tip, and $\lambda_{\mrm{f}}$ is the friction force.
\\
Impacts are possible when the contact between absorber and upper or lower cavity wall is closed.
These impacts are modeled with the \emph{Newton impact law},
\begin{equation} \label{eq:impulse}
0\leq \gamma_i^+ + e\gamma_i^-\perp \Lambda_{\mrm{u,}i}\geq 0\quad i=1,2\fp
\end{equation}
Herein, $\gamma_i= \dot g_i$ is the relative velocity, where $\gamma_i>0$ means opening contact.
$\gamma_i^+$, $\gamma_i^-$ are the velocities after and before a specific time instant, $e$ is the coefficient of restitution, and $\Lambda_{\mrm{u,}i}$ is an impulse, \ie, the integral of a force over an infinitesimal time increment.
When the contact opens with $\gamma_i>0$, no velocity jump occurs and there is no impulse.
In contrast, if the contact is closed with $\gamma_i^-<0$, the relative velocity will jump and a positive impulse $\Lambda_{\mrm{u,}i}>0$ occurs.
The velocity after the impulse, $ \gamma_i^+$, is $-e \gamma_i^-$.
For $e=1$, the normal contact interactions are purely conservative, while for $0\leq e<1$ energy is dissipated with every impulse.
\\
It can generally be stated that there is no consensus on how to model dry friction and impacts between vibrating flexible structures.
Besides the above described \emph{rigid} contact laws, \emph{compliant} contact laws are popular.
In their simplest form, they can be seen as a penalty regularization of the Signorini and Coulomb law.
Effectively, a linear spring acts during the unilateral normal contact with the wall, and a linear spring is placed in series with a Coulomb slider, see \eg \cite{Krack2017}.
To demonstrate that the findings of this study are robust with respect to a variation of the contact law, some results with compliant contact models are reported in Appendix A.
\\
Contact forces and impulses can be combined to an integral measure, the contact percussions \cite{Acary2008,Leine2013}.
Accounting for these contact percussions turns the system of ordinary differential equations \erefo{eqmbeam},\erefo{eqmabs} into measure differential equations.
In full accordance with the modal truncation of the beam model, \eref{eqmmod} has to be evaluated to determine the contact gaps and velocities.
Similarly, the contact percussions are projected onto the modal basis.

\subsection{Remark on scalability}\label{sec:scaling}
It is common practice to introduce a normalization scheme with the goal to reduce the number of model parameters.
This can facilitate transferring the results to different parameter sets.
However, we prefer to refrain from this procedure, as we believe that this would substantially reduce the clarity of the model, which does not justify the benefits of such a normalization.
\\
We would like to point out one important scaling property:
If one multiplies the gap $g_0$ by a real positive constant $\alpha\in\mathbb R$, all model equations hold if also the generalized coordinates, velocities, forces and impulses are multiplied by $\alpha$.
This can be confirmed by looking closely at \erefs{eqmbeam}-\erefo{impulse} and recognizing the linearity of the system's equations of motion (without contact) as well as the linearity of the complementarity problems describing the contact interactions.
In particular, this implies that if we know the response $\bs q_\mrm{s}(t)$, $q_\mrm{a}(t)$ for a given set of parameters, we can immediately follow that the response for $\alpha g_0$, $\alpha\mu f_{\mathrm N}$, $\alpha \hat f_{\mrm{exc}}$ is $\alpha\bs q_\mrm{s}(t)$, $\alpha q_\mrm{a}(t)$.
We will keep the gap $g_0$ fixed in this study and vary $\hat f_{\mrm{exc}}$ to determine the absorber's operating range.
Once the optimum operating point is found, the scaling property permits to design $g_0$ and $f_{\mathrm N}$ for optimal performance at any given excitation level $\hat f_{\mrm{exc}}$.

\subsection{Simulation: Time domain} \label{sec:simti}
The Moreau time-stepping scheme is used for numerical integration of the measure differential equations. 
The contact laws are enforced using an augmented Lagrangian approach in conjunction with the Gauss-Seidel relaxation method (see \eg \cite{Studer2011}).
For details on the simulation method adopted in this study we refer to~\cite{Krack2016}.
Based on a preliminary convergence study, the time step $\Delta t$ was specified in such a way that the period of the highest-frequency mode is sampled with $20$ time steps, \ie, $\Delta t = 2\pi/(20\omega_{N_\mrm{m}})$.

\subsection{Simulation: Frequency domain}
All core results in this work are based on time domain simulations.
To illustrate and interpret the periodic response regimes in \sref{responseregimes} and \sref{scattering}, the time domain simulations are supplemented by frequency domain simulations.
To this end, Harmonic Balance is used as implemented in the Matlab tool \emph{NLvib} \cite{Krack2018}.
To resolve the set-valued Signorini and Coulomb laws, the Dynamic Lagrangian approach is utilized \cite{Nacivet2003}.
It should be emphasized that this formulation does not involve the Newton impact law.
Besides the excitation frequency, the zeroth and the $20$ higher harmonics were considered to adequately resolve the contribution of off-resonant modes\footnote{For the frequency domain simulations, no more than $N_\mrm{m}=3$ modes are considered. This results in the highest natural frequency component being $\omega_3 \approx 17.5 \omega_1$ which justifies the truncation at 20 higher harmonics.}.
To overcome convergence problems, a small viscous damper $c_\mrm{a}=0.001$ had to be attached between absorber and ground, only for the frequency domain simulations.
As no significant influence on the dynamics of the system originating from the damper $c_\mrm{a}$ has been observed in any of the results obtained from time domain simulations, this slight model inconsistency was deemed appropriate.

\section{Typical response regimes and operating range}\label{sec:modquali}
In this section, we demonstrate that our model is qualified to reproduce the generally known dynamic regimes and operating ranges of impact absorbers.
This revisit might also be useful for readers not or less familiar with the dynamics of NES and impact absorbers under near-resonant forcing.
In addition, we introduce performance measures used in the original research sections \srefo{scattering}-\srefo{comparison}.
It is important to emphasize that the results presented in this section can be explained qualitatively even with a single-degree-of-freedom model of the primary structure.
We focus on the steady-state forced vibration response in the excitation frequency range around $\omega_1$, \ie, near the lowest-frequency mode of the primary structure (first bending mode).
\begin{figure}[t!]
	\centering
	\includegraphics[width=1.0\textwidth]{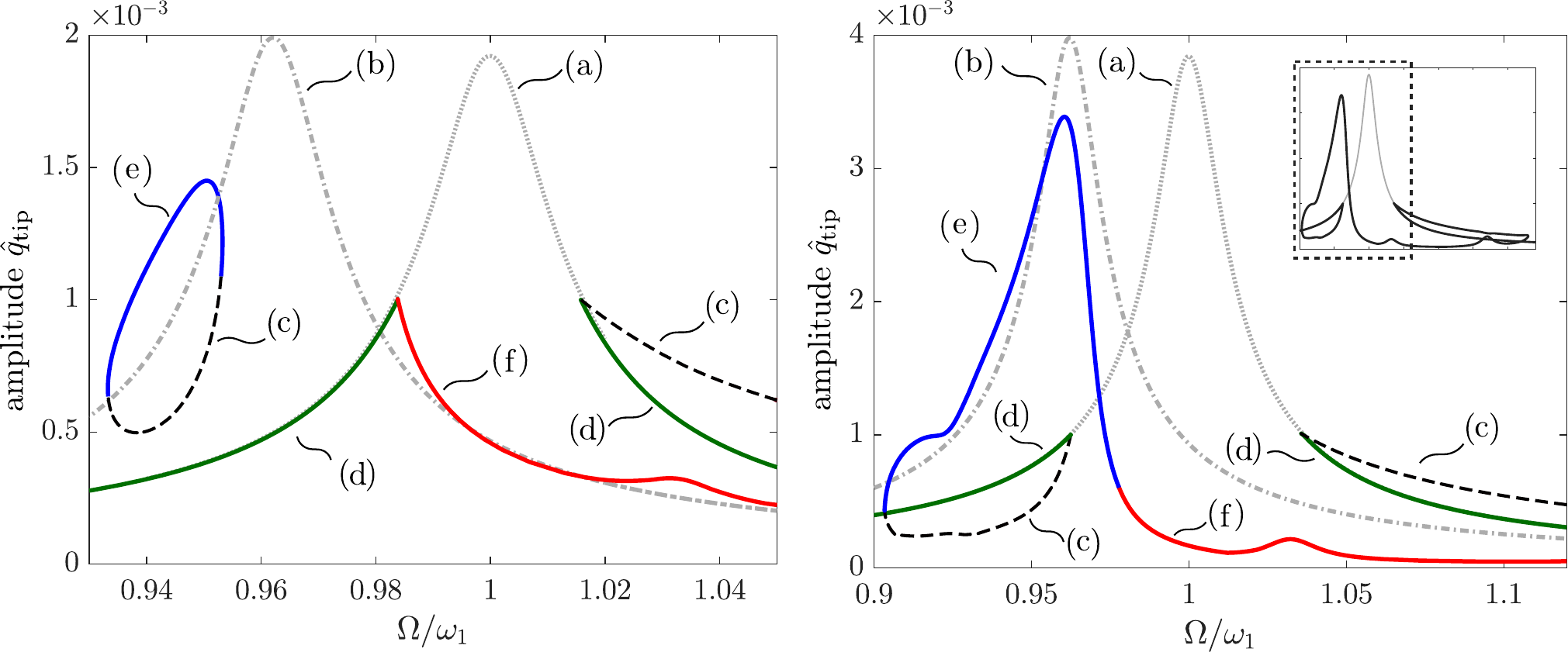}
	\caption{Typical frequency responses, left: medium level of excitation $\hat{f}_\mrm{exc}=1.5$, right: high level of excitation $\hat{f}_\mrm{exc}=3$. Parameters: $N_\mrm{m}=3$, $\mu f_\mrm{N}=0.01$.}
	\label{fig:high_low}
\end{figure}
\subsection{Typical response regimes}\label{sec:responseregimes}
\fref{high_low} shows the periodic steady-state frequency response regimes. 
The results are depicted for two representative excitation levels, a moderate (left) and a high (right) one.
The results were computed using Harmonic Balance.
The response level is here defined as the peak-to-peak/2 amplitude, $\hat{q}_\mrm{tip}$, of the periodic tip displacement $q_\mrm{tip}(t)$.
Besides the nonlinear response, the responses for two linear cases are also depicted: The beam without absorber (dotted curve (a)) and the beam with fully sticking absorber (dash-dotted curve (b)).
The resonance frequency is slightly lower in the latter case, since the absorber effectively acts as an additional mass $m_{\mathrm{a}}$ attached to the beam's tip.
\\
For a given excitation frequency and otherwise fixed parameters, there are either one, two or three periodic nonlinear responses.
Some of the responses are unstable, they will not be reached in practice.
At the turning points of the amplitude-frequency curves, saddle node bifurcations occur.
The dashed branches connecting the turning points correspond to unstable vibration states (c). We note here that the determination of asymptotic stability is performed by time domain simulations as an estimation by means of frequency domain approaches did not prove conclusive for the given system.

\subsubsection*{Transition to the linear response}
In the depicted case, the friction limit force $\mu f_{\mathrm N}$ was specified relatively small, so that the absorber is mainly sliding long before the first impact.
As a consequence, the response (d), corresponding to a purely sliding absorber without impacts, closely follows the linear one without absorber (dotted curve).
The frequency response has sharp bends at the transition from the quasi-linear response to the response with impacts.

\subsubsection*{Isolated periodic response}
For moderate excitation levels, an \emph{isola}, \ie, a closed frequency response curve occurs below the primary structure's natural frequency $\omega_1$.
At a certain excitation level, the isola merges with the main branch, to give rise to the form in \fref{high_low}-right. The inset in \fref{high_low}-right shows that the analysis indeed provides a single branch revealing a turning point to the far right of the resonance which however is outside the region of interest.
Once the branches merge, the high-level branch in blue (e), corresponding to motions with sliding absorber and impacts, can be easily reached under near-resonance forcing, \eg when one slowly sweeps through the depicted frequency range.
In this case, the resonance peak is merely shifted and only slightly decreased.
Hence, the absorber is not very effective for this excitation level, and this situation is commonly considered beyond the \emph{desired operating range}.
In contrast, if the isolated response branch occurs in some distance from the main branch, 
a strong perturbation is needed to trigger a transition from the low to the high-level response, \eg during a frequency sweep.
In \cite{Starosvetsky2008} it is suggested to suppress the high-level isola by increasing the linear viscous damping between absorber and primary structure.
If the high-level response is excluded, the absorber is quite effective. The excitation level specified for \fref{high_low}-left is near the optimal operating point of the impact absorber.
The operating ranges are further analyzed in \sref{oprange}

\subsubsection*{Strongly modulated response}
At the linear natural frequency $\omega_1$, the level of the periodic forced response is considerably reduced.
However, the depicted branches (f) of period-one responses, \ie, responses with the same fundamental frequency as the excitation, are in fact also unstable.
The stable steady-state behavior in this frequency range is a regular or chaotic \emph{strongly modulated response} (SMR) \cite{Gendelman2016}.
This regime is characterized by recurring strong changes of the amplitude envelope of the primary structure.
In this range, impacts occur in a regular fashion only during the high-amplitude part of the modulation.
This is addressed in more detail in \sref{smr}.

\subsubsection*{Almost periodic response}
The branches of high-level response in blue (e) correspond to period-one vibrations with two impacts per excitation period.
A closer look into the time domain simulations shows that the response in this range is actually only \emph{almost periodic}, it is a chaotic \emph{weakly modulated response}.
Otherwise, the nonlinear dynamics and contact interactions are very similar to the perfectly periodic response \cite{Shaw2006}.
The impacts are continuously \emph{sustained}, \ie, they occur in every excitation period, in contrast to the SMR regime.
This almost periodic regime is also referred to as \emph{chatter-type response} or \emph{constant amplitude response} \cite{Shaw2006,Pennisi2015}.

\subsubsection*{Dynamics under frequency sweeps through resonance}
The transient response under external forcing with linearly in-/decreasing frequency is depicted in \fref{jump}.
The maxima of the periodic displacement response, determined using Harmonic Balance, are also depicted.
\begin{figure}[t!]
	\centering
	\includegraphics[width=1.0\textwidth]{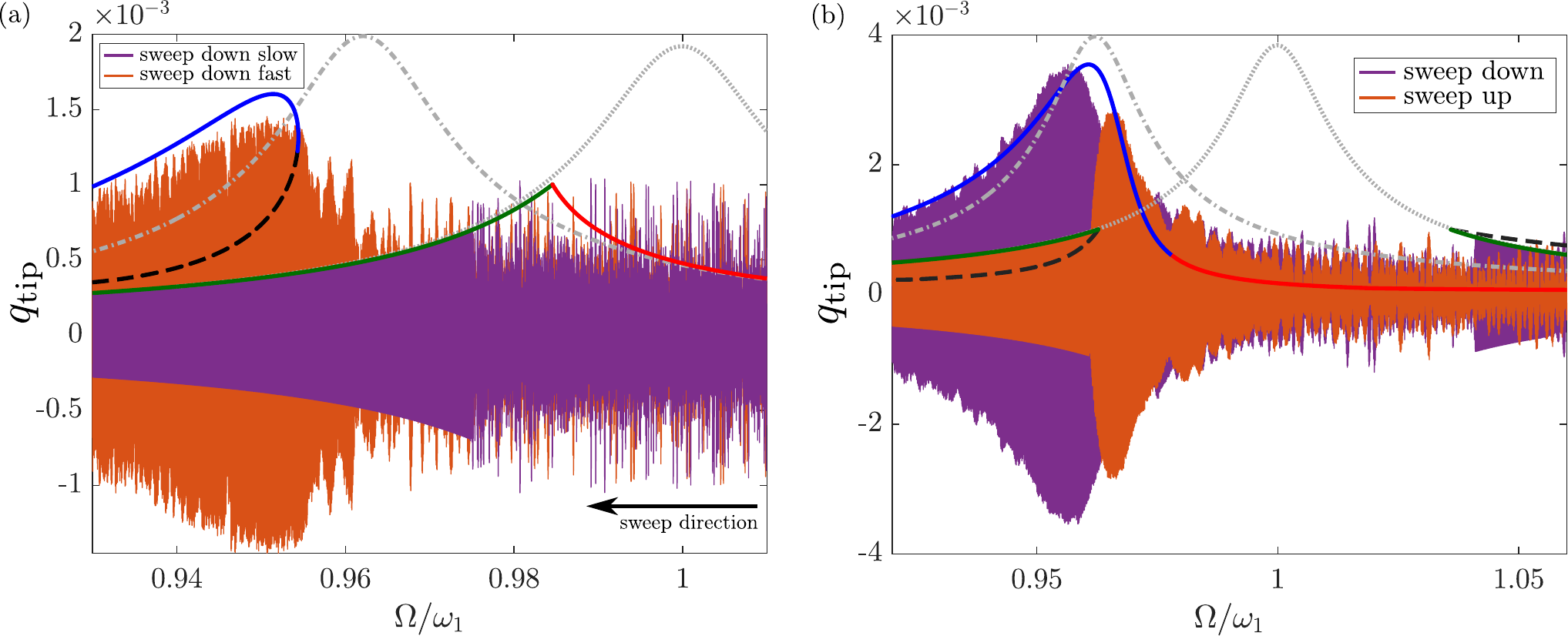}
	\caption{Response to a frequency sweep through resonance: (a) medium excitation level $\hat{f}_\mrm{exc}=1.5$, (b) high excitation level $\hat{f}_\mrm{exc}=3$. Parameters: $N_\mrm{m}=2$, $\mu f_\mrm{N}=0.025$.}
	\label{fig:jump}
\end{figure}
%
In general, it can be stated that the frequency sweep response closely follows the envelopes suggested by the stable periodic response regimes. 
This includes the high-level almost periodic steady-state response (blue curve, labeled (e) in \fref{high_low}).
Also, the SMR regime can be identified, especially in the excitation frequency range near $\Omega\approx \omega_1$.
The response maxima within the SMR regime exceed the unstable periodic regime in this frequency range (red curve, labeled (f) in \fref{high_low}), but remain well below the linear resonance.
\\
As mentioned before, the high-level isolated response regime occurring for moderate excitation levels can be reached by perturbations from the steady state low-level response.
Such a perturbation might as well be a fast transient sweep downwards through resonance as is depicted in Fig.~\ref{fig:jump}a.
Here, the sweep rate was such that the excitation frequency changes by $1\%$ in 725 pseudo periods, where a pseudo period is defined as the period $T_\mrm{p}=\frac{2\pi}{\omega_1}$.
This transition does not occur for sufficiently slow sweeps, as in the illustrated case with half the sweep rate.
This implies that the common practice of passing through resonances as fast as possible might actually be detrimental in this case.

\subsection{Repetitive targeted energy transfer during SMR}\label{sec:smr}
As observed, the absorber is particularly effective in the SMR regime.
In this subsection, different explanations for this are presented.
As mentioned before, the effects of off-resonant modes and dissipation by dry friction and impacts are addressed in detail in \srefs{scattering}-\srefo{dissipation}.
These effects are not essential to explain the effectiveness of the absorber in the SMR regime.
To separate the individual contributions to the vibration mitigation, the above mentioned effects are deliberately suppressed in this subsection.
To this end, the modal truncation order is set to $N_\mrm{m}=1$, and the contact behavior is made conservative by switching off friction with $\mu f_\mrm{N}=0$, and setting the coefficient of restitution to $e=1$.
\\
First, a typical time section of the steady-state SMR regime is determined for $\Omega=\omega_1$.
The time histories of beam tip and absorber displacements, $q_{\mathrm{tip}}(t)$ and $q_{\mathrm{a}}(t)$, are depicted in \fref{SMR}a.
As time variable, the number of pseudo periods $N_\mrm{p}=\frac{\Delta t}{T_\mrm{p}}$ is used.
At the time instants of the impacts, the relative velocity jumps and $q_{\mathrm{a}}$ has kinks.
In between the impacts, no forces act on the absorber and, hence, the mass moves with constant velocity.
During the high-amplitude time span of the SMR, the impacts occur in a sustained repetitive fashion, with about two impacts per excitation period.
In this time span, absorber and beam tip oscillate with essentially the same frequency and phase.
To show this, we computed the Hilbert transform and extracted the instantaneous phase lag $\psi_\mrm{a-s}$ between the coordinates $q_\mrm{tip}$ and $q_\mrm{a}$.
The result is depicted in \fref{SMR}b.
Indeed, $\psi_\mrm{a-s}$ approaches zero, which confirms the synchronous oscillation of absorber and beam tip.
In contrast, in the low-amplitude time span of the SMR, the phase lag assumes arbitrary values; the impacts occur in a less regular and less frequent fashion.
The regular dynamics during the high-amplitude time span can be viewed as a transient resonance phenomenon, often referred to as \emph{transient resonance capture} \cite{Vakakis2008}.
The resonance frequency ratio is here 1:1.
\begin{figure}[t!]
	\centering
	\includegraphics[width=1.0\textwidth]{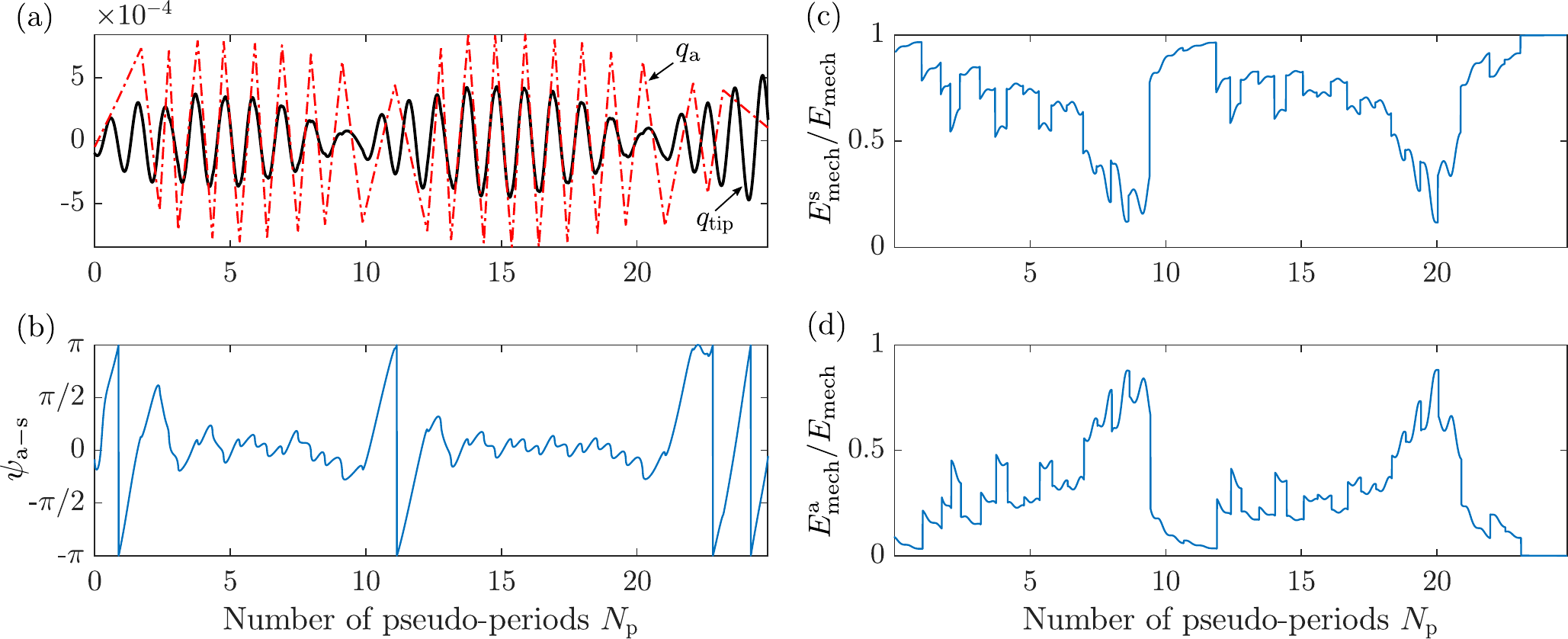}
	\caption{Dynamics in the SMR regime: (a) time history of the beam tip and absorber displacement, (b) instantaneous phase lag between beam tip and absorber, (c)-(d) energy distribution within the system. Parameters: off-resonant modes suppressed ($N_\mrm{m}=1$), conservative contact ($e=1, \mu f_\mrm{N}=0$), $\hat{f}_\mrm{exc}=1.7$, $\Omega=\omega_1$.}
	\label{fig:SMR}
\end{figure}
\\
The transient resonance capture facilitates the energy transfer between absorber and primary structure.
To illustrate this, we define the following measures corresponding to the instantaneous mechanical energy distributed among the considered modes of the primary structure and the absorber:
\begin{equation} \label{eq:Einst}
\begin{aligned}
E_\mrm{mech}^\mrm{s} &= \sum\limits_{i=1}^{N_\mrm{m}} E_\mrm{mech}^{\mrm{mod}_i} = \sum\limits_{i=1}^{N_\mrm{m}} \left(E_\mrm{pot}^{\mrm{mod}_i} + E_\mrm{kin}^{\mrm{mod}_i}\right) = 
\sum\limits_{i=1}^{N_\mrm{m}} \frac12 \omega_i^2\eta_i^2 + \frac12 {\dot \eta}_i^2\fk\\
E_\mrm{mech}^\mrm{a} &= \dfrac{1}{2}m_\mrm{a}\dot{q}_\mrm{a}^2\fp
\end{aligned}
\end{equation}
Note that the absorber can only store kinetic energy.
The energy quantities were computed for the simulated time section and are depicted in \fref{SMR}c-d.
As mentioned above, we set $N_\mrm{m}=1$ here, so that $E_\mrm{mech}^\mrm{s}=E_\mrm{mech}^{\mrm{mod}_1}$, \ie, all mechanical energy of the primary structure is contained in the first bending mode.
Apparently, the energy transfer is so effective that temporarily up to 90\% of the total instantaneous energy is localized in the absorber, here at $N_\mrm{p}\approx8$ and $N_\mrm{p}\approx20$.
Note that the time span of effective energy transfer to the absorber coincides with the time span of transient resonance capture, as indicated by $\psi_\mrm{a-s}\approx 0$ here.
If the loading was impulsive and the absorber could directly dissipate energy, this energy transfer would be targeted only towards the absorber and irreversible.
However, since the loading is harmonic, the targeted energy transfer occurs in a repetitive form.
It is important to note that the LVTA also relies on the targeted energy transfer, triggered by the linear resonance of the secondary structure.
Hence, it can only occur near the resonance frequency of the LVTA.
In contrast, no such preferential frequency exists in the case of an essential nonlinear attachment of the absorber.
The repetitive targeted energy transfer during SMR can take place at any frequency, which is the key to the broad-band effect of the nonlinear absorber.
It is well known that the SMR regime is generally the desired operating regime of NES: Besides facilitating effective targeted energy transfer, it can be shown that it has a relatively large basin of attraction \cite{Vakakis2008}.
This is particularly important if high-level periodic response regimes co-exist.
For analytical predictions of SMR regimes we refer to \cite{Vakakis2008}, for the specific case of non-smooth NES to \cite{Gourc2015,Lamarque2011}.
\\
A completely different explanation for the high effectiveness of impact absorbers under near-resonant forcing is given in \cite{Hartung2016}.
Here, the vibration mitigation concept is termed \emph{impulse mistuning}, and it is explained by frequency detuning triggered by impacts:
When the vibration level of the primary structure is low, there are no impacts and the absorber has no effect.
The vibrations will then grow due to the linear resonance.
Once the vibration level is sufficiently large, impacts will occur.
This changes the system's effective resonance frequency.
During the unilateral contact phases, the absorber effectively adds mass to the primary structure, thus the effective resonance frequency is decreased.
As a consequence of the impact dynamics, the external forcing with the natural frequency of the primary structure (without absorber) is out of resonance.
Thus, the vibrations will decay and the process repeats itself.
This can be viewed as an alternative way of explaining both the existence of SMR and the passive decrease of the near-resonant vibration level as compared to the linear case without absorber.

\subsection{Operating ranges}\label{sec:oprange}
As mentioned before, the overall vibration mitigation performance and the critical dynamic regime of an impact absorber depends on the excitation level.
To investigate this, a suitable definition of the resonant response level is needed.
The main difficulty here is that various steady-state responses may co-exist at any given excitation frequency.
This could be addressed by a probabilistic approach, see \eg \cite{Vakakis2008}: A statistical distribution of the initial conditions is assumed and a statistical distribution of the response level can be determined using conventional Monte Carlo Simulation.
In this work, we use a more practical approach: We define the resonant response level as the \emph{maximum} response level reached during a moderately slow \emph{frequency sweep through resonance}.
It should be stressed that periodic excitations are typical for rotating machinery, which occur in almost all energy and transport systems.
As the rotor speed is varied, resonance situations may occur.
The proposed definition of the resonant response level is thus a measure for the maximum response level reached during a \emph{resonance passage}, \ie, a transient crossing of a resonance situation under variation of the rotor speed.
Of course, due to the well-known hysteresis effects, the maximum response level may significantly depend on the sweep direction (\cf, for instance, \fref{jump}b).
Thus, both up and down sweeps are evaluated, and the maximum of both is taken.
A minimum excitation frequency of $0.9\omega_1$ and maximum $1.06\omega_1$ were considered, and the initial displacements and velocities are set to zero.
The frequency limits are well within the (quasi-)linear response regime.
This means, in particular, that the absorber is initially centered between upper and lower cavity walls.
A crucial aspect of the proposed definition is the sweep rate.
Throughout the remaining part of the article, the sweep rate was selected so that the instantaneous excitation frequency changes by $1\%$ every 100 pseudo periods.
This is much faster than the very slow sweep rates used in \fref{jump}.
This makes jumps to the possibly detached high-level response regime more likely, and is regarded as relatively conservative.
\\
It should be remarked that a strategy used at times is to assess the performance of impact absorbers only for a single frequency, \eg $\Omega=\omega_1$.
As we will see later, this leads to a substantial underestimation of the critical response level near resonance.
One could argue that the performance improvement for a single frequency might still be of technical relevance for systems where the excitation frequency is invariant and known.
However, in such a situation, one could achieve the same effect simply with a slightly modified linear design, which is most likely more robust and easier to implement.
\\
The second difficulty is to define a suitable measure for the response level itself.
In periodic response regimes, taking the peak-to-peak amplitude is common practice.
In weakly or strongly modulated response regimes, this might not be a fair or robust measure.
Instead, we used the following procedure:
First, determine the time instant $t_\mrm{max}$ of the maximum magnitude of the considered coordinate, in this study the beam's tip displacement $q_{\mathrm{tip}}$,
\e{ t_{\mrm{max}} = \operatorname{argmax} \left|q_{\mathrm{tip}}(t)\right|\fk}{tmax}
during the entire resonance passage.
We then define the \emph{resonant response level} $\bar{q}_\mrm{tip}$ as root mean square of $q_{\mathrm{tip}}$ in the time span of $\pm 10$ pseudo periods around $t_{\mrm{max}}$ (windowed RMS),
\begin{equation}
\bar{q}_\mrm{tip}=\sqrt{\dfrac{1}{t_{\mathrm e}-t_{\mathrm s}}\int\limits_{t_{\mathrm s}}^{t_{\mathrm e}}q^2_\mrm{tip}(t)\,\mrm{d}t},\quad \text{with} \quad t_{\mathrm s}=t_\mrm{max}-10T_\mrm{p},\quad t_{\mathrm e}=t_\mrm{max}+10T_\mrm{p}\fp \label{eq:barqtip}
\end{equation}
%
The length of the time span will obviously influence $\bar{q}_\mrm{tip}$.
We note that \eg doubling the considered time span to $\pm 20$ pseudo periods reduces the overall level of $\bar{q}_\mrm{tip}$, but all of our conclusions remain invariant.
\begin{figure}[b!]
	\centering
	\includegraphics[width=1\textwidth]{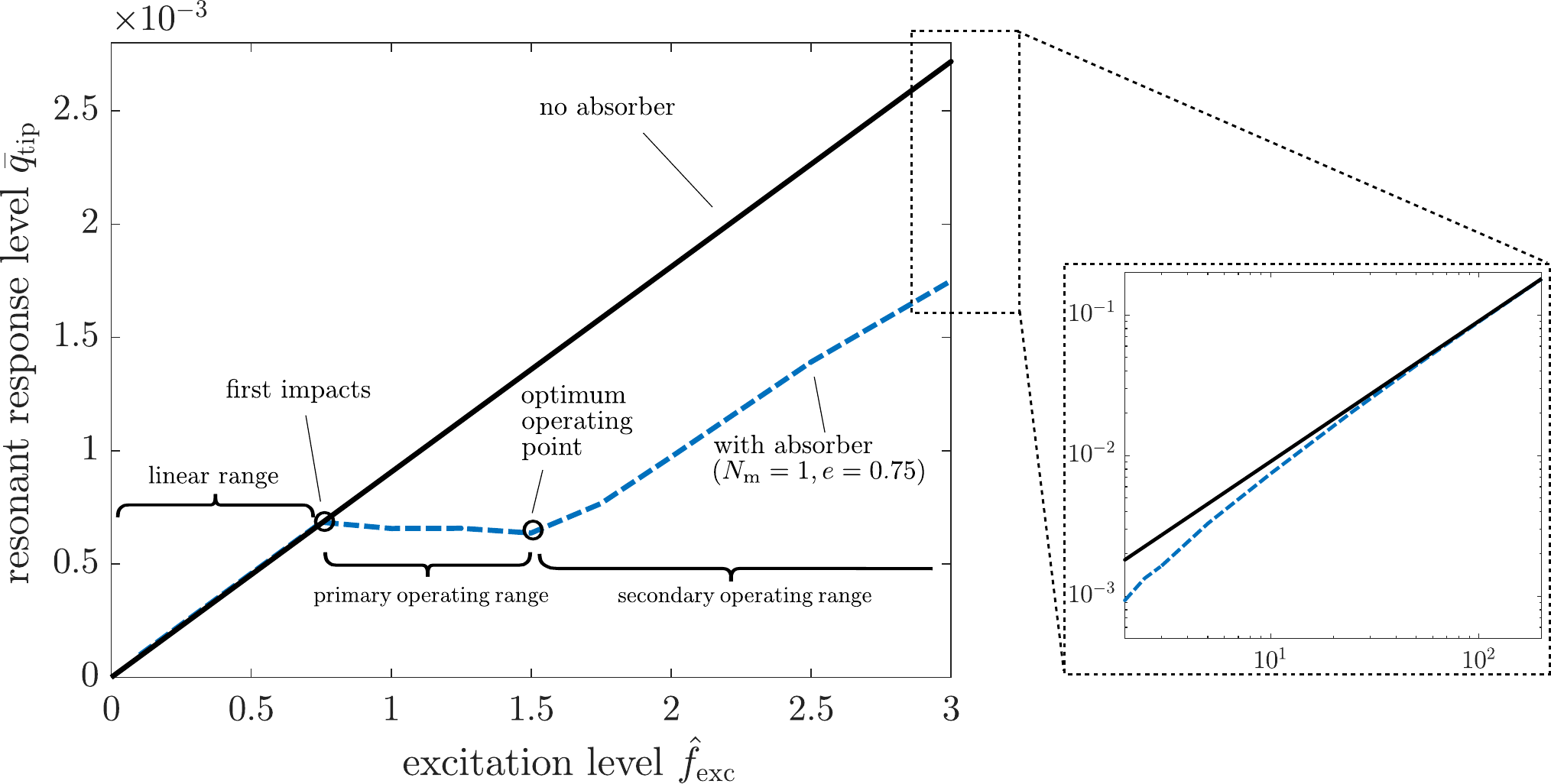}
	\caption{Typical absorber performance curve for a system with and without an impact absorber.}
	\label{fig:oprange} 
\end{figure}
%
%
\begin{table}[b!]
	\centering
	\caption{Operating ranges with associated contact interactions and response type}
	\begin{tabular}{|c|c|c|}
		\hline
        operating range                     & contact interactions                                      & response\\
		\hline
        linear range                             & no impacts                                                       & harmonic\\
        primary operating range     & impacts $\leftrightarrow$ no impacts    & strongly modulated\\
        secondary operating range& sustained impacts                                          & weakly modulated/almost periodic\\
		\hline
	\end{tabular} \label{tab:ranges}
\end{table}
\\
In \fref{oprange}, a typical \emph{absorber performance curve} is shown, which depicts the characteristic relationship between resonant response level, $\bar{q}_\mrm{tip}$, and excitation level, $\hat{f}_\mrm{exc}$.
It can be divided into three distinctive operating ranges which are also compactly summarized in \tref{ranges}:
\begin{enumerate}
\item
\emph{linear range:} At sufficiently low excitation levels, no impacts occur, leading to linear behavior.
\\
\item
\emph{primary operating range:}
Upon the first impacts, the primary operating range starts.
Here the critical response is strongly modulated such that time spans with sustained, repetitive impacts alternate with time spans with infrequent impacts.
In this range, the resonant response level exhibits \emph{saturation} with respect to the excitation level (almost horizontal section of performance curve).
\\
\item
\emph{secondary operating range:}
At a certain excitation level, the almost periodic high-level response is reached, either due to transient jumps to the isolated steady-state response branch (\cf \fref{jump}), or due to the merging of this branch with the main branch beyond a certain excitation level (\cf \frefs{high_low} and \frefo{jump}).
Here, impacts are sustained, with two or more impacts per pseudo period.
From this transition on, the dependence on the excitation level seems quasi-linear.
For extremely high response levels, the clearance $g_0$ becomes negligible as compared with the system's vibration level.
The absorber then practically acts as an additional mass fixed to the beam's tip.
Thus, the resonant response characteristic approaches this linear case for very high $\hat{f}_\mrm{exc}$ as is apparent from the inset in \fref{oprange}.
It should be remarked that the performance curve for this linear case is virtually identical to that without absorber.
\end{enumerate}
%
Obviously, the \emph{optimum operating point} is achieved at the transition point between (almost) periodic response (secondary operating range) and strongly modulated response (primary operating range).
This was also shown theoretically and experimentally for impact absorbers in \cite{Li2016}.
As discussed in \sref{scaling}, the scaling property can be used to design the clearance between absorber and cavity walls to achieve best performance for a given excitation level.

\section{Effect of off-resonant modes}\label{sec:scattering}
In \sref{oprange}, we approximated the beam dynamics only by the resonant mode, $N_{\mathrm m} =1$, and thus deliberately suppressed the effect of off-resonant modes.
Recall that off-resonant modes were not given much attention in the literature so far, when studying NES under periodic excitation.
Primary structures with multiple modes are commonly considered, but mainly to demonstrate that the NES is effective for more than one resonance.
In comparison to low-order polynomial type stiffness nonlinearities, impacts greatly facilitate the energy transfer to off-resonant modes.
In this section, we show that off-resonant modes contribute to the performance of an impact absorber to a considerable extent.
We also demonstrate that in the case of conservative contact behavior between absorber and primary structure, the absorber is predicted to be noneffective if the primary structure is approximated by only a single mode.
We show that \emph{for conservative contact behavior}, only the energy transfer to and dissipation by \emph{off-resonant vibration modes explains the high effectiveness of the impact absorber}.
\begin{figure}[t!]
	\centering
	\includegraphics[width=1\textwidth]{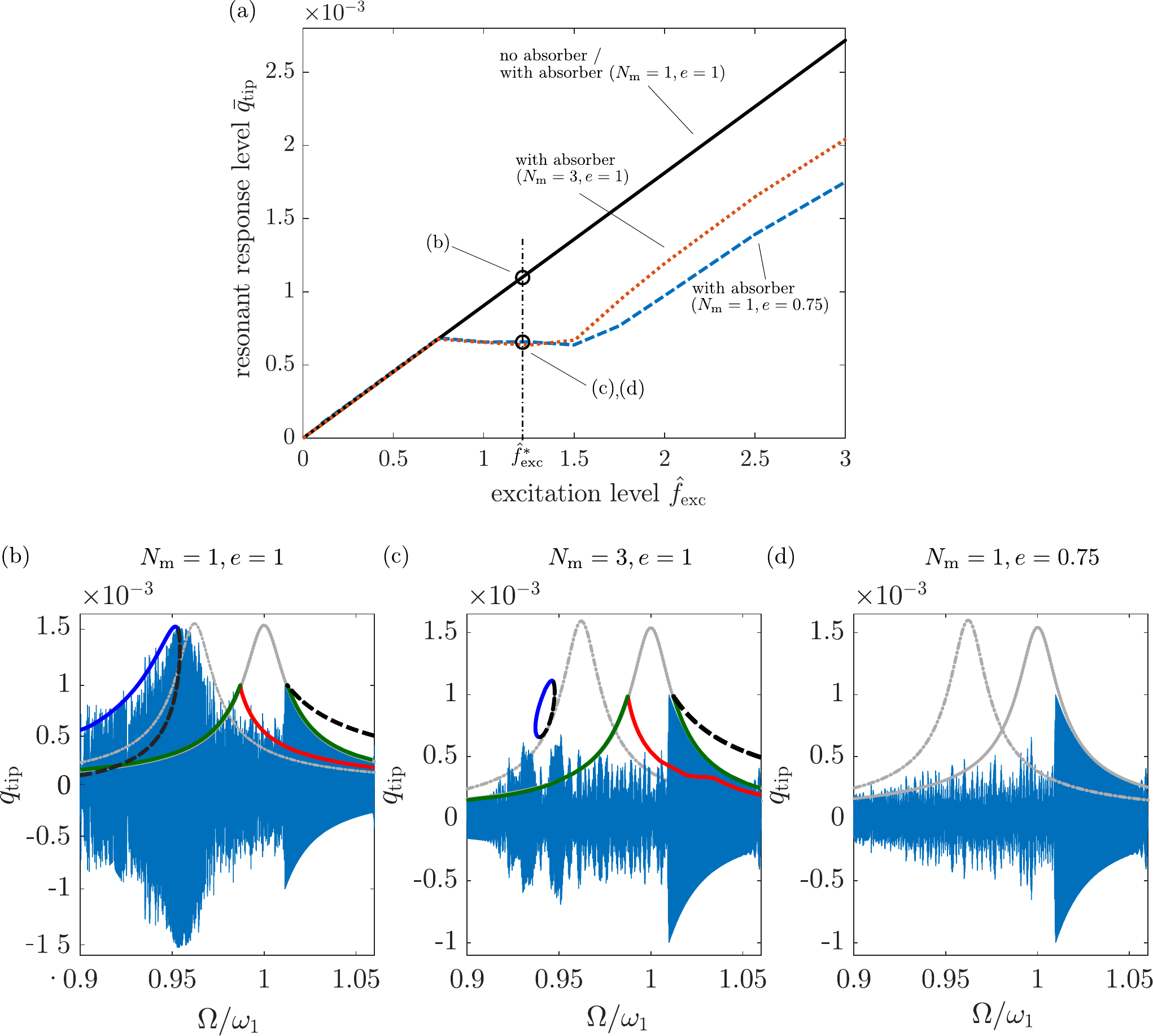}
	\caption{Absorber performance for both conservative and non-conservative contact behavior as well as $N_\mrm{m}\geq 1$: (a) performance curves, (b)-(d) frequency sweep response for a representative excitation level $f_\mrm{exc}^\ast$.}
	\label{fig:oprangeTWO} 
\end{figure}

\subsection{Effectiveness for conservative contact behavior}
In \fref{oprangeTWO}a the absorber performance curve is depicted for \emph{conservative contact behavior}, \ie, with $e=1$, $\mu f_\mrm{N}=0$, for two cases $N_{\mathrm m}=1$ and $N_{\mathrm m}=3$.
Remarkably, if the contact is conservative and $N_{\mathrm m}=1$, the resonant response level coincides with that of the system without absorber.
Hence, the absorber is noneffective in this case.
The reason for this can be seen in the frequency response in \fref{oprangeTWO}b.
Apparently, the vibrations are effectively suppressed very close to the natural frequency of the system without absorber, \ie, at $\Omega\approx \omega_1$.
Here, the repetitive targeted energy transfer during the SMR causes a considerable decrease of the vibration level as compared to the case without absorber.
However, during the critical down sweep through resonance, the high-level response regime is reached.
The maximum response level is located at $\Omega/\omega_1\approx 0.95$ in this case.
The response level is very close to that of the system without absorber (\cf \fref{oprangeTWO}a and b).
This makes sense as the absorber here mainly acts as an additional mass, while additional damping by impacts or off-resonant modes is suppressed.
Previous works have often focussed on the response level precisely at $\Omega= \omega_1$, which would suggest high effectiveness of impact absorbers even for $N_{\mathrm m}=1$ and $e=1$.
The results in \fref{oprangeTWO} clearly demonstrate that the critical response peak is not reduced in this case, but merely shifted to a few percent lower frequencies.
Thus, we conclude that impact absorbers are ineffective for suppressing forced vibrations near resonance for $N_{\mathrm m}=1$ and $e=1$.
\\
It is interesting to see that for dissipative impacts with \eg $e=0.75$, the high-level response regime is not reached for otherwise identical parameters, \cf \fref{oprangeTWO}d.
The corresponding regime has either ceased to exist, or it is simply not reached during the transient resonance passages. As was discussed in \sref{simti} the Newton impact law with $e<1$ cannot be accounted for in the Dynamic Lagrangian approach and, hence, no comparative results obtained from frequency domain simulations are not depicted in \fref{oprangeTWO}d.
The maximum response level is then reached at the occurrence of first impacts, \ie, at the transition from the linear case without impacts to the nonlinear one.
Since for the transient sweeps, the absorber is initially placed at the cavity center, $q_{\mathrm a}=0$, this transition is reached when the linear response reaches $\left|q_\mrm{tip}\right|=g_0$ which is again regarded as a conservative approach.
A similar effect can also be achieved for non-dissipative impacts, $e=1$, but with $N_\mrm{m}>1$, \cf \fref{oprangeTWO}c.
The steady-state frequency response results indicate that the energy transfer to and dissipation by off-resonant modes reduce the high-level response branch.
Also, this response regime occurs on an isolated branch, which is well separated for $N_\mrm{m}=3$ at the depicted excitation level $f_\mrm{exc}^*$.
This makes the undesired jump to the high-level branch less likely during resonance passages.\\
The sweeps in \fref{oprangeTWO}b and c reveal that the dynamics are not necessarily attracted to the stable periodic response (purely sliding absorber in green) that is predicted by the frequency domain simulations. This stems from the transient nature of the sweep excitation as well as the low damping. For example, the SMR-regime persists at the green-red transition close to $\Omega/\omega_1=1$ because the system is perturbed by the occurring impacts when reaching this point. For the excitation level and parameters chosen in the presented cases the dynamics are only actually attracted to the stable periodic response in this frequency range for very low sweep rates as seen from \fref{jump}a in purple.

\subsection{Energy transfer to and dissipation by off-resonant modes}
In the following, we analyze in more detail the dynamic interactions between resonant and off-resonant modes, and the resulting effects on the absorber performance.
We start by considering only two modes of the primary structure, by setting $N_\mrm{m}=2$, thus, suppressing the higher vibration modes.
We will later show that the dynamics within the SMR regime is rather insensitive to the number of considered vibration modes.
Hence, we focus for now on the secondary operating range, where the critical response level is reached in the high-level almost periodic regime with sustained impacts.
To this end, the excitation level is set to $\hat{f}_\mrm{exc}=3$, which is located well within the secondary operating range (\cf \fref{oprangeTWO}a).
The contact behavior is made conservative by setting $e=1$, $\mu f_\mrm{N}=0$.
For the given beam model, the frequency ratio between the first two bending modes is constant with $\omega_2/\omega_1 = 6.3$.
In spite of this, it is interesting to explore the effect of the frequency ratio.
This was investigated by artificially varying $\omega_2$ in the modal formulation of the beam model.
The results are depicted in \fref{varom}.
The nonlinear resonant response level is normalized by the corresponding value of the linear system without absorber, $\bar{q}_\mrm{tip}^\mrm{no\,abs}$.
\begin{figure}[b!]
	\centering
	\includegraphics[width=1.0\textwidth]{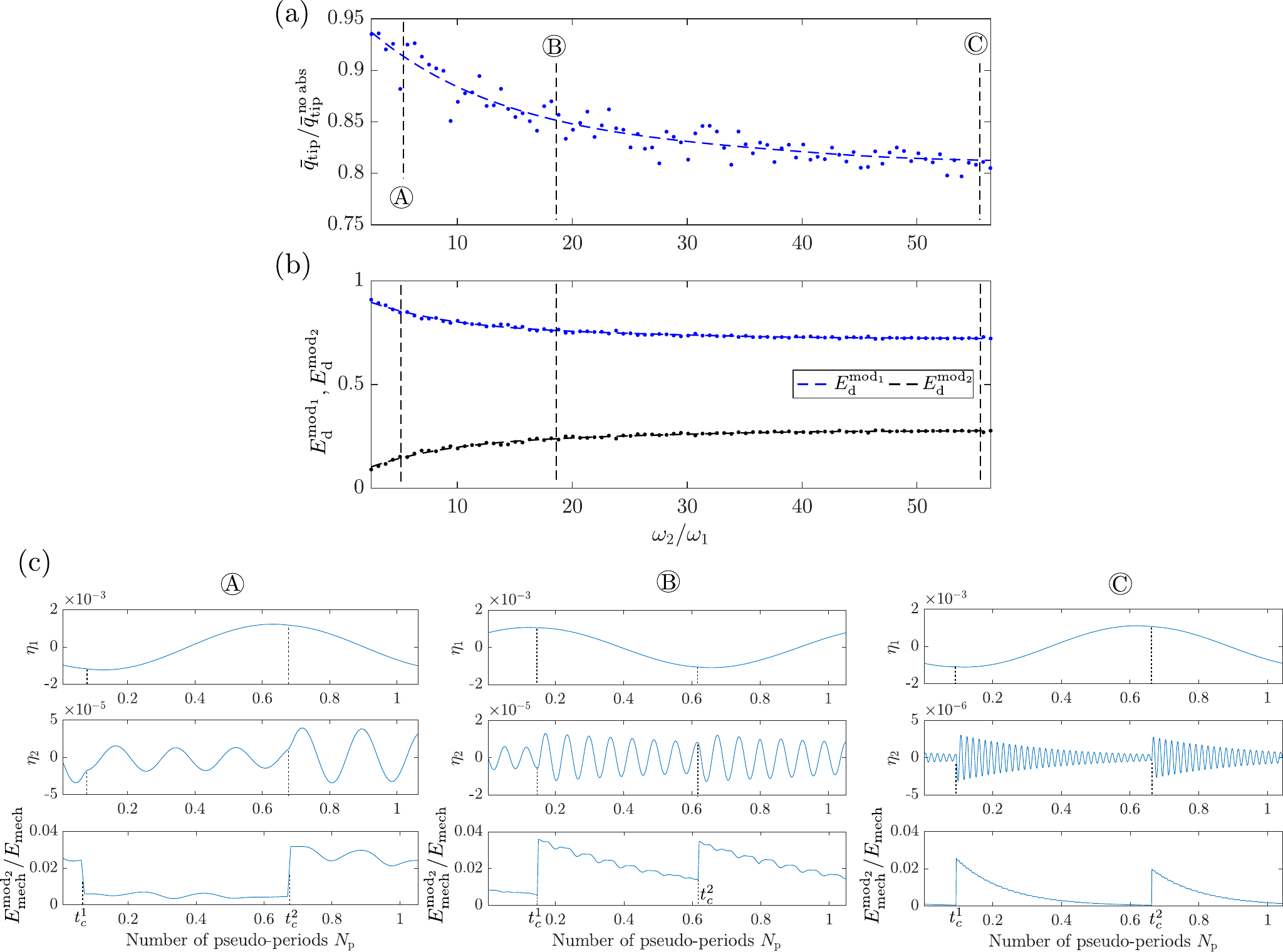}
	\caption{Modal energy distribution vs. frequency of the off-resonant mode $\omega_2$, $\hat{f}_\mrm{exc}=3$ (secondary operating range); $D_1=D_2 = 1\%, N_\mrm{m}=2$.}
	\label{fig:varom}
\end{figure}
\begin{figure}[b!]
	\centering
	\includegraphics[width=1.0\textwidth]{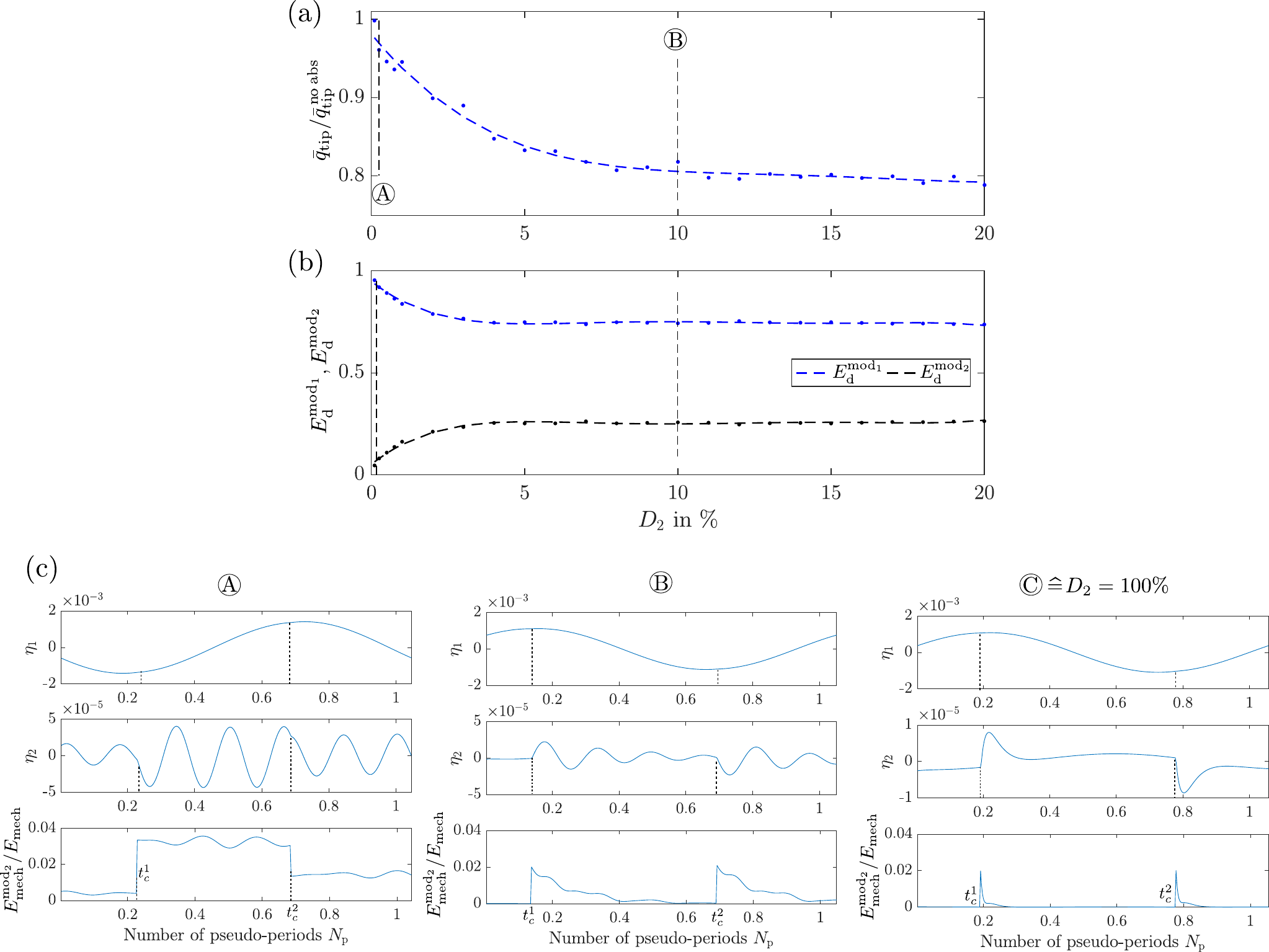}
	\caption{Modal energy distribution vs. damping of the off-resonant mode $D_2$, $\hat{f}_\mrm{exc}=3$ (secondary operating range); $D_1=1\%$, $\omega_2 = 6.3 \omega_1, N_\mrm{m}=2$.}
	\label{fig:vard}
\end{figure}
The energy dissipated in each mode, $E_\mrm{d}^{\mrm{mod}_i}$, is computed as
\e{E_\mrm{d}^{\mrm{mod}_i} = \int\limits_{t_\mathrm{s}}^{t_{\mathrm e}} 2D_i\omega_i \dot \eta_i^2\dd t\fk}{edmi}
and it is given by relating it to the sum of all modal contributions, \ie, $E_\mrm{d}^{\mrm{mod}_1}+E_\mrm{d}^{\mrm{mod}_2}$ here (\cf \fref{varom}b).
\\
Per pseudo period, two or more impacts occur in the secondary operating range. In the depicted intervals, two impacts occur at the time instants $t_\mrm{c}^1$, $t_\mrm{c}^2$.
At each impact, a certain amount of mechanical energy is transferred from the resonant mode, coordinate $\eta_1$, to the off-resonant higher vibration mode, coordinate $\eta_2$.
The amount is $\approx 2-3\%$ per impact in the depicted cases (\cf \fref{varom}c, bottom figures).
In the time span between two impacts, the system behaves linearly, and only the modal damping causes dissipation.
Since the higher mode is off-resonant, its dynamics are virtually unaffected by the external forcing and, thus, dominated by their natural dynamics.
For the same modal damping ratio, the energy fraction dissipated per period $T_2=2\pi/\omega_2$ is invariant.
However, as $\omega_2$ increases, the period $T_2$ shortens and the modal energy $E_{\mathrm{mech}}^{\mrm{mod}_2}$ dissipates faster.
As a consequence, more energy dissipates between two impacts.
This explains the improvement of the absorber performance as the frequency ratio increases (\cf \fref{varom}a).
For impulsive excitation, this phenomenon is also referred to as \emph{modal energy scattering}.
In the following, we will refer to this shortly as \emph{dispersion}, as the effect mainly relies on the distribution of energy input to one wave form (resonant mode) to other wave forms (off-resonant modes).
For large frequency ratios $\omega_2/\omega_1$, virtually all energy transferred to the off-resonant mode is dissipated until the next impact.
A further increase of the frequency ratio then cannot lead to more energy dissipation, and therefore leads to no significant change in the overall dynamics.
Thus, the \emph{damping capacity of the higher mode is limited}.
The bottleneck is the amount of energy that can be transferred to the higher mode by an impact.
This explains the saturation effect for higher frequency ratios, where the resonant response level and the modal contributions to energy dissipation are seen to approach asymptotic values.
In this case, the resonant response level is improved by $\approx 20\%$.
Hence, it can be stated that \emph{dispersion contributes considerably to the absorber performance in the secondary operating range}.
\\
A similar effect can be achieved for fixed frequency ratio $\omega_2/\omega_1 = 6.3$ but increased damping ratio $D_2$.
This is illustrated in \fref{vard}.
For sufficiently high damping ratio, again virtually all energy input to the second mode is dissipated before the next impact occurs.
\\
Finally, we analyze the effect of additional off-resonant modes by successively increasing the modal truncation order $N_{\mrm{m}}$ as presented in \fref{scat}.
The damping ratios are all set to their nominal value of $D_i=1\%$, and the natural frequencies $\omega_i$ are determined by the beam model.
Contact is further assumed as conservative, \ie, $e=1$, $\mu f_\mrm{N}=0$.
Two excitation levels are considered, $\hat{f}_\mrm{exc}=5$ and $\hat{f}_\mrm{exc}=1.5$.
For nominal parameters, $\hat{f}_\mrm{exc}=5$ is located well inside the secondary operating range, whereas $\hat{f}_\mrm{exc}=1.5$ is located inside the primary operating range.
However, as explained above, the operating range actually depends on the modal truncation order.
For $N_\mrm{m}=1$ and conservative contact, the primary operating range practically does not exist.
For $\hat{f}_\mrm{exc}=1.5$, we depict not the maximum response level over the entire resonance passage, but only within the SMR regime, \ie, in a narrow frequency range around $\Omega/\omega_1= 1$.
This permits to analyze the effect of off-resonant modes for the different dynamic regimes separately (rather than for the different operating ranges).
\begin{figure}[b!]
	\centering
	\includegraphics[width=1.0\textwidth]{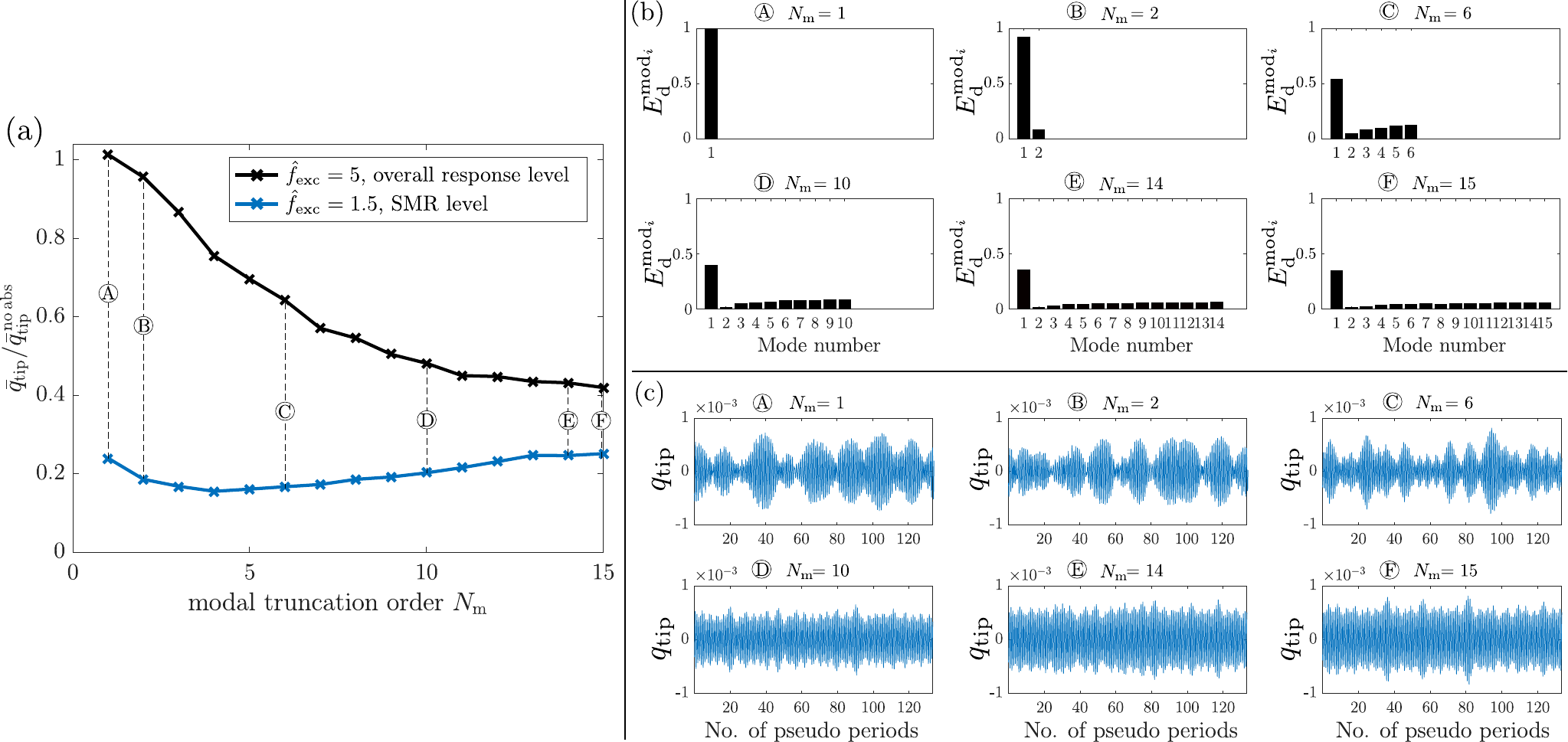}
	\caption{Contribution of off-resonant modes to the absorber performance, for $\hat{f}_\mrm{exc}=5$ and for $\hat{f}_\mrm{exc}=1.5$; (a) resonant response level vs. modal truncation order $N_\mrm{m}$, (b) distribution of dissipated energy among the modes for $\hat{f}_\mrm{exc}=5$, (c) SMR dynamics for $\hat{f}_\mrm{exc}=1.5$.}
	\label{fig:scat}
\end{figure}
%
\\
For $\hat{f}_\mrm{exc}=5$, the overall resonant response level with $N_\mrm{m}=1$ is virtually identical to that without absorber (\cf also \fref{oprangeTWO}b).
Hence, no reduction is achieved and $\bar{q}_\mrm{tip}/\bar{q}_\mrm{tip}^{\mathrm{no\,abs}} \approx 1$.
The reduction of the overall response level for $N_\mrm{m}>1$ indicates that dispersion causes a considerable performance enhancement in the secondary operating range.
Still, the relative majority of the energy is confined and dissipated in the resonant mode (\cf \fref{scat}b).
The energy dissipated by each off-resonant mode is much smaller, and, as explained before, strictly limited.
However, the sum of all off-resonant contributions to the dissipated energy is significant and causes an overall resonant response level reduction by $\approx 60\%$ compared to $N_\mrm{m}=1$ here.
Tracking the instantaneous energy distribution among all modes reveals that only as of approximately mode 6 the frequency is high enough to allow for a (on average) complete dissipation of energy transferred to the respective mode before the next impact. Furthermore, the amount of energy transferred per impact to the modes 6 to $N_\mrm{m}$ is approximately the same which can be explained by the Fourier transform of the impulse in the form of a Dirac delta distribution. This explains the nearly uniform distribution of the energy that is damped by these modes (\cf \fref{scat}b).
\fref{scat}a suggests a stabilization of the behavior with respect to $N_\mrm{m}$.
We would like to emphasize, however, that the results are here depicted for the case of purely conservative contact behavior.
The quantitative influence of off-resonant modes should be further analyzed in a laboratory test.
\\
For $\hat{f}_\mrm{exc}=1.5$, the focus is placed on the SMR regime.
To this end, the SMR level $\bar{q}_\mrm{tip}$ is measured only in the range of $N_\mrm{p}=130$ pseudo periods centered around $\Omega=\omega_1$.
This corresponds to the frequency range of $0.9935<\Omega/\omega_1<1.0065$.
The associated time signals of $q_\mrm{tip}(t)$ are depicted in \fref{scat}c.
In contrast to the overall response level, the response level in the SMR range is reduced down to $\approx 20\%$ even for $N_\mrm{m}=1$, \ie, without dispersion to off-resonant modes as was outlined in \sref{smr}.
Recall at this point that the response level was defined as root mean square value and not the peak response of the time signals in \fref{scat}c.
While the strong modulation can be clearly seen for small $N_\mrm{m}$, the modulation becomes less obvious for large $N_\mrm{m}$. This decrease in degree of modulation with increasing dispersion is an interesting aspect that is left for further investigation and explanation in future work. 
Apart from this, the off-resonant modes have no considerable influence on the SMR level.

\section{Dissipation by dry friction and impacts}\label{sec:dissipation}
The investigations in the previous section showed that high effectiveness of impact absorbers can be explained by energy transfer and dissipation by off-resonant modes (dispersion), even for the case of conservative contact behavior.
In this section, we analyze the effect of dissipative dry friction, $\mu f_\mrm{N}>0$, and dissipative impacts, $e<1$.
To account for dispersion as well, the modal truncation order is set to $N_\mrm{m}=10$ in accordance with the modal convergence analysis in \sref{scattering} (\cf \fref{scat}a).
The absorber performance curve is depicted in \fref{dissinf}a for different friction and impact behavior.
The cases without absorber (solid black) and conservative contact behavior (dash-dotted gray) are shown again for orientation.
\begin{figure}[t!]
	\centering
	\includegraphics[width=1.0\textwidth]{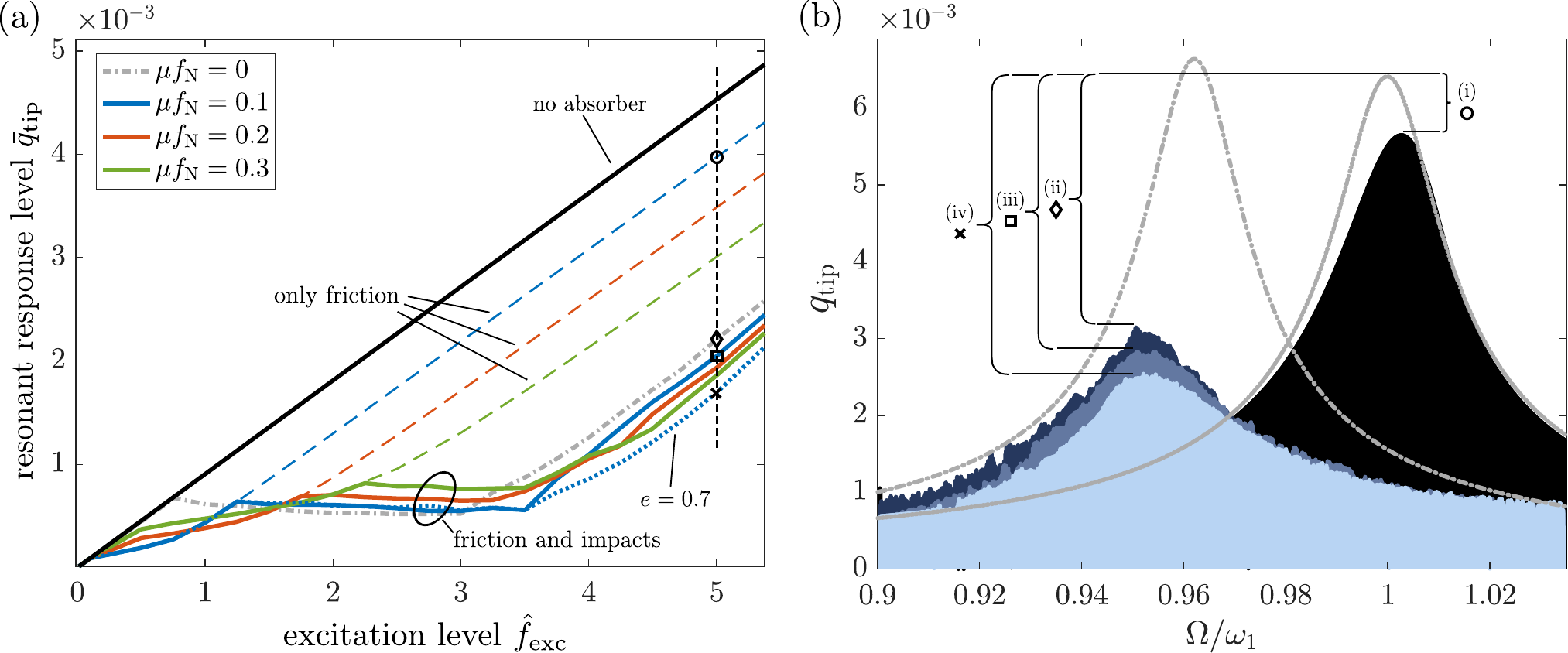}
	\caption{Effect of dissipation by dry friction and impacts: (a) absorber performance curve, (b) frequency sweep response for $\hat{f}_\mrm{exc}=5$.}
	\label{fig:dissinf}
\end{figure}
\\
For sufficiently large gap $g_0$, no impacts occur and the absorber acts as a pure friction damper (dashed lines in \fref{dissinf}a).
Similar to the impact absorber, the dry friction damper has (quasi-)linear limit cases, one at low excitation levels (only sticking) and one at high excitation levels (mainly sliding friction), and a maximum damping performance for some intermediate value.
The location of the optimum operating point depends on the friction limit force $\mu f_\mrm{N}$.
For a higher friction limit force, the optimum operating point is reached for a higher excitation level.
\\
When the system is sufficiently loaded that the absorber can slide, and $g_0$ is sufficiently small, impacts occur.
As a consequence, the solid curves (with impacts) depart from the corresponding dashed curves (only friction, without impacts).
The point of first impacts is reached at higher excitation levels under larger friction limit forces $\mu f_\mrm{N}$.
For small $\mu f_\mrm{N}$, the optimum friction damper operating point is reached well before the optimum impact absorber operating point.
The primary operating range for small friction limit force, $\mu f_\mrm{N}=0.1$, is on almost the same level as that for the frictionless case.
However, the range appears to be shifted along the $\hat{f}_\mrm{exc}$ axis.
Higher friction limit forces lead to worse performance in the primary operating range of the impact absorber.
The opposite tends to be the case for the secondary operating range ($\hat{f}_\mrm{exc}>3.5$).
The performance in the secondary operating range, where the critical response regime is characterized by sustained impacts, can be further improved by additional dissipation during impacts, \eg with $e=0.7$.
In contrast, the resonant response level in the primary operating range cannot be further reduced by this.
\\
\fref{dissinf}b shows the response to downward frequency sweeps through resonance for an excitation level of $\hat{f}_\mrm{exc}=5$.
It summarizes the effects contributing to the vibration mitigation in the secondary operating range of the impact absorber:
(i) friction damping ( \APLcirc{} ), (ii) modal energy scattering ($\lozenge$), (iii) friction damping + scattering (\Square), and (iv) friction damping + scattering + dissipative impacts ($\times$).
\\
The main takeaway message from this section is that the performance of the considered impact absorber is relatively robust against dry friction. This is beneficial in case the normal load is unknown or cannot be controlled.
Dry friction has only a comparatively small effect on the response level in the main operating ranges.
However, the shift of the optimum operating point to higher excitation levels should be taken into account in the design of the absorber.

\section{Effectiveness for multiple resonances in comparison to LTVA and friction damper} \label{sec:comparison}
In this section, we assess the performance of the impact absorber in comparison to a pure friction damper and a conventional LTVA.
As mentioned before, a major drawback of the LTVA is that it can be tuned to only a single resonance, while it remains virtually ineffective for the remaining resonances.
To effectively suppress two resonances, one would thus need two LTVAs and so on.
This can soon become an infeasible design task.
In contrast, it is well known that the same impact absorber can be effective for multiple resonances.
For the comparative assessment of the different absorber/damper concepts, we considered the primary resonances with the first and second mode of the base structure, \ie, the frequency ranges around $\Omega=\omega_1$ and $\Omega=\omega_2$.
For fairness, the same absorber/damper mass is used for all vibration mitigation concepts.
For the impact absorber, dissipation by both dry friction and impacts is considered.
\begin{figure}[t!]
	\centering	
	\includegraphics[width=1.0\textwidth]{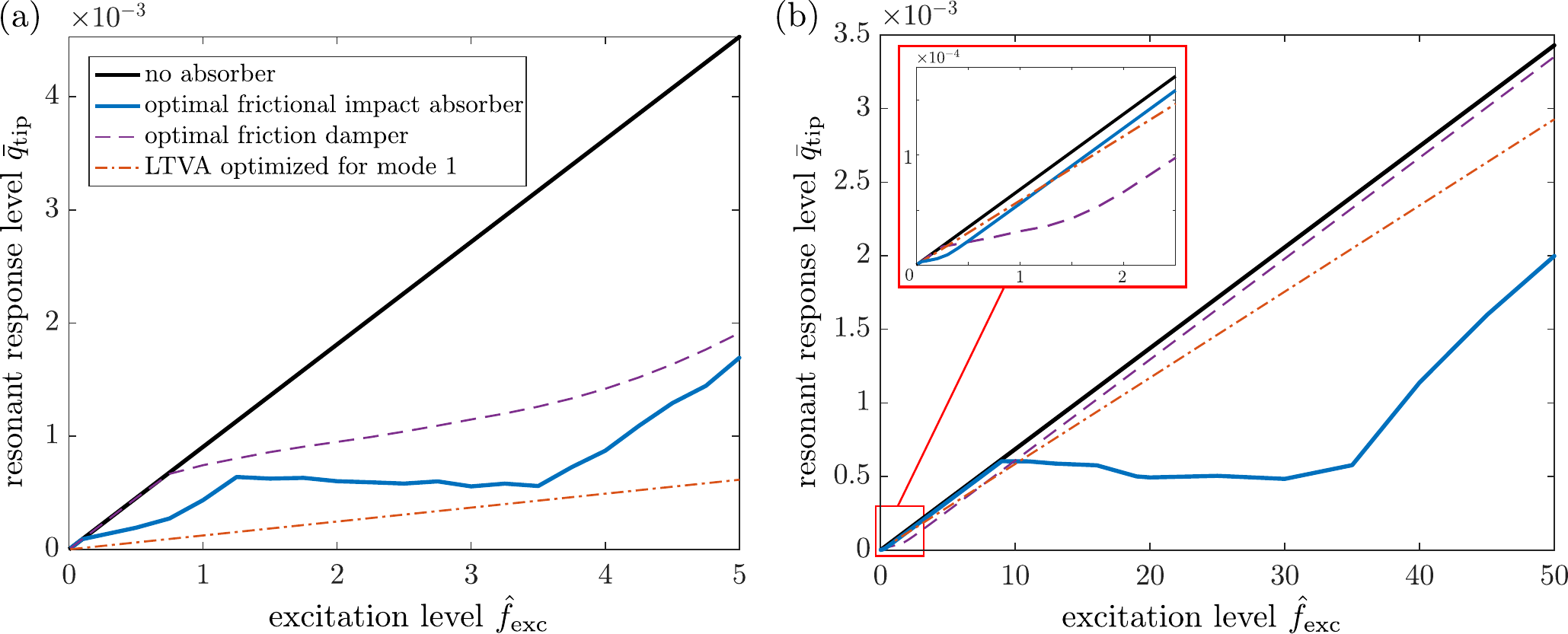}
	\caption{Performance curves for different absorber/damper concepts near primary resonance of (a) the first and (b) the second mode. Parameters: optimal impact absorber (--): $\mu f_\mrm{N}=0.1,e=0.7$; optimal friction damper (- -): $\mu f_\mrm{N}=0.5$; LTVA (-$\cdotp$-) stiffness $k_\mrm{a,s}=701.8$, damper $c_\mrm{a,s}=0.728$.}
	\label{fig:m12}
\end{figure}
\subsubsection*{Design of the dry friction damper}
The behavior of the pure dry friction damper can be easily described with the model in \sref{modsim} by excluding impacts.
This can be achieved by setting the gap $g_0$ between absorber and cavity wall to a sufficiently large value.
It is well known that friction dampers can be designed for optimal performance at a given excitation level.
Here, we optimized the performance to $\hat{f}_\mrm{exc}=3.5$, which is also the optimum operating point of the impact absorber.
To this end, we adjusted the friction limit force to $\mu f_\mrm{N}=0.5$.

\subsubsection*{Design of the LTVA}
Concerning the LTVA, the linear spring and viscous damper of the absorber attachment are tuned to the first vibration mode of the primary structure.
To this end, den-Hartog's equal peak rule~\cite{DenHartog1928} is used.
This leads to a minimal overall response level near the first resonance.
The resulting LTVA spring stiffness is $k_\mrm{a,s}=701.8$ and the viscous damper constant is $c_\mrm{a,s}=0.728$.

\subsection{Comparative performance assessment}
The performance curves for the different absorber/damper concepts are depicted in \fref{m12}.
As expected, the LTVA outperforms both the impact absorber and the friction damper near the resonance to which it was tuned.
Near its optimum operating point, the impact absorber performs almost as good as the LTVA.
At the resonance with the second mode, \ie, near $\Omega=\omega_2$\footnote{For the second resonance the sweep rate was again chosen so that the excitation frequency changes by 1\% every 100 respective pseudo periods $T_\mrm{p}=\frac{2\pi}{\omega_2}$.} the same LTVA has almost no effect.
In contrast, the impact absorber shows its typical operating characteristic and proves its high effectiveness for both resonances.
If the parameters of the impact absorber are not changed, the optimum operating point is at $\hat{f}_\mrm{exc}\approx 35$ for the second resonance, as compared to $\hat{f}_\mrm{exc}\approx 3.5$ for the first resonance. The reason for this is that for the second resonance the amplitudes at the beam tip reach critical values only for higher levels of excitation.
Hence, it is not ensured that the impact absorber performs optimal for both resonances at given excitation levels.
As mentioned before, the gap $g_0$ can be designed to achieve optimal performance at a given excitation level for a given resonance.
The results imply, however, that this setting generally leads to suboptimal performance for other resonances (at corresponding excitation levels).
To further improve the performance of an impact absorber, it might thus be reasonable to include additional geometric variables into the design problem, such as the location of the cavity within the primary structure, see \eg \cite{Zulli2014,Masri1973}.
If the design space permits this, the primary operating range could effectively be broadened by incorporating multiple impact absorbers with individual gaps \cite{Li2017}.
\\
Another important assertion from \fref{m12} is that the impact absorber performs, in a wide range, much better than a friction damper with the same mass.
For the second resonance, it can be seen that the friction damper performs worse even than the detuned LTVA in a wide range of excitation levels.
The inset in \fref{m12}b shows that the optimum operating range of the friction damper here occurs for much lower excitation levels.
In this range, it in fact outperforms the LTVA and the impact absorber for the considered parameter settings.
As the performance curves of impact absorber and friction damper are qualitatively similar (linear, primary and secondary operating ranges), the above described difficulty to achieve optimal performance for multiple resonances and given excitation levels applies to both concepts.
However, the impact absorber performs considerably better than the friction damper with the same mass.

\section{Conclusions} \label{sec:conc}
In this work, we thoroughly investigated how impact absorbers (or vibro-impact nonlinear energy sinks) mitigate resonant vibrations.
Our basic model captures
(a) strongly nonlinear unilateral interactions between absorber and primary structure,
(b) potentially dissipative impacts, and
(c) off-resonant modes.
In addition, our model accounts for inevitable dry friction between absorber and primary structure.
We found that our model is qualified to reproduce the well-known dynamic regimes and operating ranges of single-degree-of-freedom primary structures with impact absorbers.
\\
We propose a practically oriented absorber performance measure for structures under harmonic forcing: the maximum response level reached during a frequency sweep through resonance.
Using this performance measure, we demonstrated that impact absorbers are ineffective under conservative contact conditions and if no energy is transferred to higher modes of the primary structure.
The reason for this is that an almost periodic high-level response regime is reached during the sweep.
The associated response is of about the same level as that of the resonance of the system without absorber, the critical frequency is merely shifted by a few percent.
This makes impact absorbers practically useless under the described conditions, in spite of the considerably reduced vibration level in the strongly modulated response regime.
However, we also demonstrated that the picture changes completely when the energy transfer to and dissipation by off-resonant modes, \ie, the dispersion is taken into account.
Then, the impact absorber is highly effective, even under conservative contact conditions.
We believe that this is of utmost technical relevance, as the immediate dissipation by impacts is inevitably associated with local plastic deformation and, thus, most likely limits the service life of absorber or primary structure.
\\
We also demonstrated that inevitable dry friction between absorber and primary structure does neither impede nor significantly improve the performance of impact absorbers.
This is also in line with our conclusion that a pure dry friction damper is much less effective than a impact absorber of the same mass.
While the perfectly tuned conventional linear vibration absorber could not be outperformed in its designated frequency range near a particular resonance, an impact absorber has the outstanding capability to effectively suppress multiple resonances.
We further confirmed that impact absorbers have a wide, yet limited range of excitation levels for which they operate effectively.
The optimal operating point of the impact absorber can be reached for a given excitation level by exploiting the scalability property outline in this work and designing the gap accordingly.

\section*{Acknowledgments}
The authors are grateful to MTU Aero Engines AG for providing the financial support for this work and for giving permission to publish it.

\appendix
\setcounter{figure}{0}
\setcounter{table}{0}
\section{Comparison of elastic and rigid contact models}
Besides the rigid contact models used in this work, it is common to use compliant, \eg piecewise linear contact models to describe impact absorbers, see \eg \cite{Lamarque2011}.
In this appendix, we compare results from both rigid and compliant formulations, in order to demonstrate the robustness of the findings in this work with respect to the contact model.
To this end, the rigid Signorini law is replaced by a linear-elastic unilateral spring,
\begin{equation}
\lambda_{\mrm{u},i}=\begin{cases}
k_\mrm{n}g_i &\quad  g_i>0,\; \text{contact}\\
0 &\quad g_i\leq0,\; \text{no contact}
\end{cases}\fk \label{eq:elSignorini}
\end{equation}
with constant contact stiffness $k_\mrm{n}$.
The rigid Coulomb law is replaced by its elastic variant, which can be expressed by the differential law
\begin{equation}
\dd \lambda_{\mrm{f}}=\begin{cases}
k_\mrm{t}  \dd g &\quad \text{if}~~ \left| \lambda_{\mrm{f}} + k_\mrm{t}  \dd g\right|\leq\mu f_\mrm{N}~~\text{(sticking)}\\
0 & \quad \text{otherwise} ~~\text{(sliding)}
\end{cases},\label{eq:elCoulomb}
\end{equation}
with constant contact stiffness $k_\mrm{t}$ and relative displacement $g=q_\mrm{a}-q_\mrm{tip}$.
This corresponds to placing a spring $k_\mrm{t}$ in series with a Coulomb slider element.
The compliant contact models in \erefs{elSignorini}-\erefo{elCoulomb} can be interpreted as penalty regularizations of the associated rigid laws.
However, they are also commonly used to capture compliance neglected in the structural model, stemming, for instance, from surface roughness \cite{Krack2017}.
Since the compliant contact laws are single-valued, standard numerical integration methods can be used for the simulation.
We used a Runge-Kutta method.
The time step size was specified to satisfy the following two conditions:
(1) $\Delta t \leq 2\pi/(20\omega_{N_\mrm{m}})$, which ensures that the period of the highest-frequency mode is resolved with at least 20 samples.
(2) $\Delta t \leq \pi\sqrt{m_\mrm{a}/k_\mrm{n}}/20$, which ensures that the simplified contact duration \cite{Brogliato1999} is also resolved with at least 20 samples.
\\
\fref{conv}a illustrates the effect of the increasing contact stiffness $k_\mrm{n}$ on the resonant response level for $\hat{f}_\mrm{exc}=4$ for a frictionless impact absorber ($\mu f_\mrm{N}=0$). The contact stiffness values are normalized with respect to the static stiffness of the cantilevered beam, $k_\mrm{eff}=\frac{3EI}{l^3}$.
As the unilateral spring does not dissipate energy, for consistency, the coefficient of restitution was set to $e=1$ in the rigid model of the impact absorber when computing the reference value $\bar{q}_\mrm{tip}^\mrm{rigid}$.
A very similar converging behavior was encountered when performing the same analysis for a pure friction damper with increasing contact stiffness $k_\mrm{t}$ which is however not presented herein.\\
To further assess the influence of the compliant attachment, the performance curve presented earlier in \fref{oprangeTWO}a for the rigid model with $N_\mrm{m}=3, e=1$ was reproduced with the compliant model for two different values of $k_\mrm{n}$.
\fref{conv}b shows that for sufficiently high contact stiffness, the results of the compliant model approach those predicted by the rigid model for all considered levels of excitation.
From this we conclude that the findings derived in this study, are not only valid in the limit case of rigid, but also for slightly compliant contact behavior.
Of more practical relevance in the end is, of course, the performance in the physical test.
\begin{figure}[t!]
	\centering
	\includegraphics[width=1.0\textwidth]{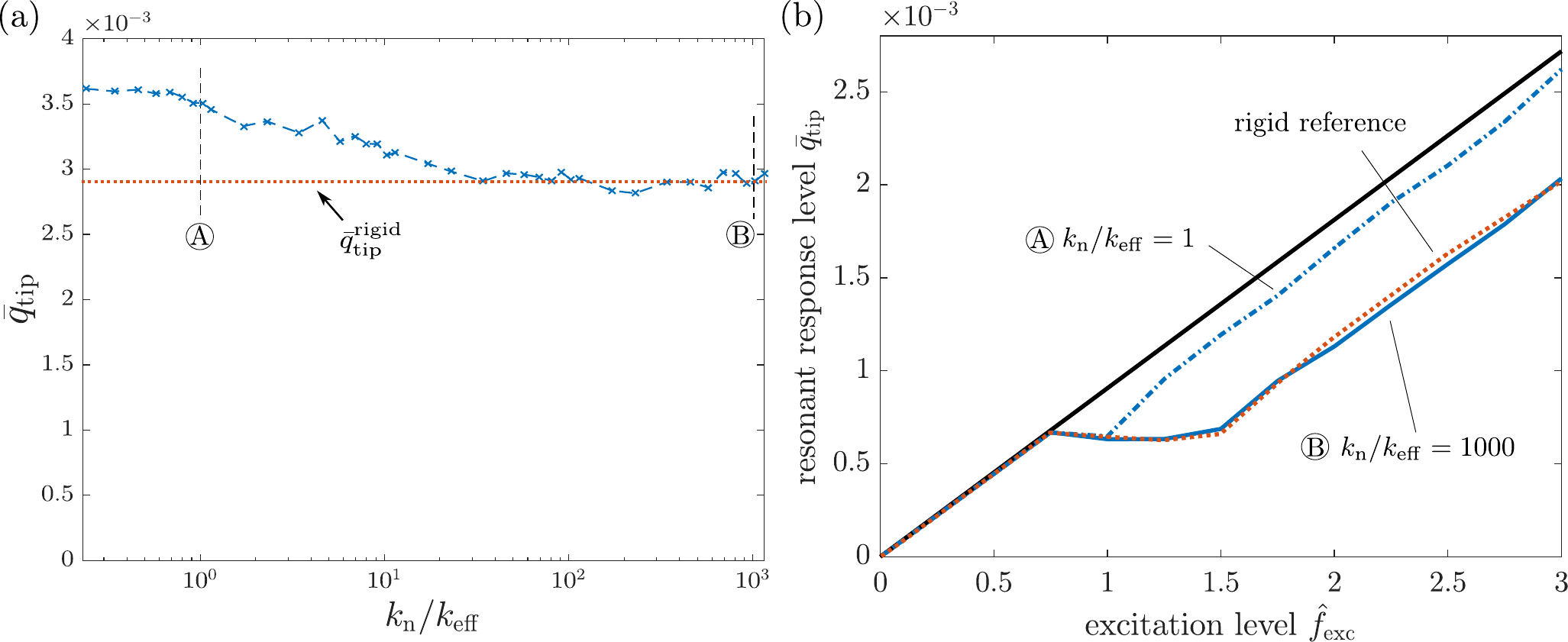}
	\caption{Effect of compliant contact: (a) resonant response level for increasing contact stiffness at single excitation level $\hat{f}_\mrm{exc}=4$, (b) performance curve for low and high contact stiffness in reference to rigid model.}
	\label{fig:conv}
\end{figure}
%
%
%

\section*{References}

\bibliography{p01}

\end{document}